\documentclass[11pt]{article}

\usepackage{amsmath,mathrsfs,amsbsy,amsthm}
\usepackage{cite}
\usepackage{graphicx,epsfig}
\usepackage{latexsym,amssymb}
\usepackage{epsf}
\usepackage{ifpdf,lineno}
\usepackage{color}
\usepackage{geometry}

\begin{document}
	
	\title{{\bf       Faster-Than-Light Solitons  in $1+1$ Dimensions}}
	
\author{M. Mohammadi$^{1}$\thanks{Corresponding Author. } \\
	{\small \texttt{physmohammadi@pgu.ac.ir}}\and
	N. Riazi$^{2}$ \\{\small \texttt{n\underline~riazi@sbu.ac.ir}}\and
	M. H. Dehghani$^{1}$ \\{\small \texttt{Mhd.pgu@gmail.com}}}
	\date{{\em{$^1$Physics Department, Persian Gulf University, Bushehr 75169, Iran.\\
$^2$ Department of Physics, Shahid Beheshti University, Evin, Tehran 19839, Iran.}}}
	\maketitle

\begin{abstract}

The existence  of a faster-than-light particle is  in direct opposition to Einstein's relativity and the principle of causality. However,  we show that the theory of classical relativistic fields   is not inherently inconsistent with the existence of the  faster-than-light particle-like  soliton solutions in $1+1$ dimensions. We introduce two extended KG models ($k$-fields)  that  lead to a zero-energy  and a nonzero-energy   stable particle-like solution with faster-than light speeds, respectively.

\end{abstract}


	\section{Introduction}\label{sec1}

The typical and classical conception  assumes  that a particle is a stable entity  that may be found at any arbitrary velocity. In the context of the  relativistic classical field theory, such  stable particle-like solutions with localized energy density functions are called solitons \footnote{According to some well-known references such as  \cite{rajarama}, a solitary wave solution is a soliton if it reappears  without any distortion after collisions. The  stability is just  a necessary condition for a solitary wave solution to be a soliton. However, in this paper,  we only accept the stability condition for the definition of a soliton solution.} \cite{rajarama,Das,lamb,TS}. In many respects,  they resemble the real stable particles. For example, they satisfy the same well-known relativistic energy-momentum relations of  special relativity and their dimension  would contract  in the direction of motion  according to the Lorentz contraction law. There are many works on relativistic solitons and solitary wave solutions, among which one can mention the kink (anti-kink)  solutions  of the real nonlinear Klein-Gordon (KG) systems in $1+1$ dimensions \cite{phi41,phi45,OV,Kink1,Kink2,Kink3,GMS,Kink5,Kink6,Kink7,Kink8,JRM1,JRM2,Kink10,Kink11,Kink12,Kink13,ana1,MR1,MR3,MR4},  the Q-ball solutions of the complex nonlinear KG systems \cite{waz,Vak3,Vak4,Vak5,Vak6,Lee3,Scoleman,R1,R2,R3,R4,R5,R6,RM,R7,R8},  the Skyrme's model  \cite{TS,SKrme,SKrme2,SKrme3,SKrme4} of  baryons,  and 't Hooft Polyakov's model
 which yields   monopole  soliton solutions \cite{rajarama,TS,toft,pol,MKP,TOF,CTOPO}.

Solitary wave solutions and solitons can be divided into two groups, topological and non-topological, depending on how they are at the boundaries. The topological solitons do not have the same behavior at far distances (boundaries) and are inevitably stable. Apart from Q-balls, the other cases mentioned above are topological solitons.  However, the non-topological solitary wave solutions have the same boundary behaviors and are not necessarily stable.  If we do not restrict ourselves to the relativistic field systems, the study of non-topological soliton solutions in several branches of physics and mathematics has been of great interest, for example, one can mention \cite{N1,N2,N3,N4,N5,N6,N7,N8,N9,N10,N11,N13}.

According to the  special theory of  relativity,   the motion of any matter  particle  is restricted to be at speeds less than the speed of light. However, for first time, the hypothetical faster-than-light (FTL) particles concept, was proposed  by Gerald Feinberg who coined the term  tachyons \cite{GF,GF2} and defined them  as the quanta of  a special relativistic   quantum field theory  with imaginary mass. The complex speeds open up another possibility  to build a theory with hypothetical particles at FTL speeds \cite{GF3}. There are some notable  works on this matter,  which can be  useful for the interested reader \cite{FTL1,FTL2}. For the hypothetical FTL  particles, the main implication  is the violation of causality, which is accepted as an obvious principle in physics. Thus, FTL particles may be discussed in theory and mathematics,  but in the real world, the existence of such particles  contradicts the accepted  axioms. It is well known that in contrast to group velocity, phase velocities can easily exceed the speed of light.


In   classical  relativistic field theory with  solitons solutions,
 there have not been introduced a special system so far, which yields a solitary wave or soliton solution at FTL speeds.  In this paper, we show  mathematically how a classical relativistic  field theory can lead to a non-topological soliton solution at FTL speeds in $1+1$ dimensions. Here, the speed of light is ordinarily  a limiting speed for the motion of the soliton  solution, that is  it   cannot   move   at  speeds less than the speed of light.
In fact, two extended KG models will be introduced which yields  zero and non-zero energy   FTL soliton solutions, respectively.  These models can be considered as  toy mathematical models   to show  that   the theory of relativistic classical fields  is not  inherently inconsistent with the existence of  FTL particle-like solutions.


The extended KG systems  or the so-called $k$-fields, have Lagrangian densities which are not linear in the kinetic scalar terms ${\cal S}_{ij}={\cal S}_{ji}=\partial_{\mu}\phi_{i}\partial^{\mu}\phi_{j}$  \cite{Baz1,Adam,Bab,MMQ1,MMQ2,MMQ4,PH3,MMQ3}. Such Lagrangian densities  can be also called non-canonical  Lagrangian (NCL) densities   \cite{SH,ZZ,Rami0,Rami,Rami2,Rami3}. The solitary wave and soliton solutions of such systems are known as defect structures \cite{Baz1,Adam,Bab}.     There are many works which deal  with such systems with defect structures (e.g.,  domain walls, vortices and monopoles), among which one can mention  \cite{Baz1,Adam,Bab,Vash,Mahzon}.
In cosmology, the models  with $k$-fields have become especially popular. They are suggested in the context of inflation leading to   $k$-inflation \cite{Armend,Chiba,Armend2}, or they   are used  for describing dark energy and dark matter \cite{Armend3,R2,cos1,cos2,cos3}.



The organization of this paper is as follows: In Section \ref{sec2}, we will introduce a standard nonlinear KG system with an unstable FTL solitary wave solution.  In Section \ref{sec3},   zero energy solitary wave solutions are  introduced first, then  an extended KG will be introduced with an energetically  stable zero-energy soliton solution at FTL speeds. In section 4, by combining these two models, we will introduce a new system that leads to a stable FTL solitary wave solution with nonzero-energy. The last section is devoted to  summary and conclusion.

\section{An unstable solitary wave solution  at FTL speeds}\label{sec2}

In the standard relativistic  theory of the classical fields, it is common to start  with  a proper Lagrangian density and then try to find its   solitary   wave  solutions.    There is another approach where one can first consider a special  proposed solitary   wave solution and then try to find a proper  Lagrangian density for it \cite{MR1}. A solitary wave solution is a special solution that  has a localized  energy density function. For example, based on the second approach, for a  real scalar field  $\varphi$, we can consider   a nonlinear KG Lagrangian density,
\begin{equation} \label{L}
{\cal L}=\partial^\mu \varphi\partial_\mu \varphi- U(\varphi)=\dot{\varphi}-\varphi'-U(\varphi),
\end{equation}
which is assumed to have  a special   localized  Gaussian  solution at rest in the following form:
\begin{equation} \label{ssw}
\varphi_{o}=\exp(-x^2).
\end{equation}
Here $U(\varphi)$ is called the field  potential and  should be determined  in such a way that Eq.~(\ref{ssw}) becomes  a  special  solution of the Lagrangian density (\ref{L}). Note that, in Eq.~(\ref{L}), the dot (prime) indicates the time (space) derivative, and for the sake of simplicity, throughout the paper, we assume the speed of light to be equal to one.
In fact, Eq.~(\ref{ssw}) is considered to be a special solution of the dynamical equation, which results from  the Lagrangian density (\ref{L}):
\begin{equation} \label{de}
 \Box \varphi =\frac{\partial^2\varphi}{\partial
  t^2}- \frac{\partial^2\varphi}{\partial
  x^2}=-\frac{1}{2}\frac{d U}{d\varphi}.
  \end{equation}
 Hence, for the proposed static solution (\ref{ssw}), the dynamical equation (\ref{de}) is reduced to
\begin{equation} \label{s}
\frac{d^2\varphi_{o}}{d
  x^2}=\frac{1}{2}\frac{d U(\varphi_{o})}{d\varphi_{o}}.
\end{equation}
Moreover, from Eq.~(\ref{ssw}), one can invert  $x$ as a function of $\varphi_{o}$ , i.e. $x=\pm\sqrt{-\ln|\varphi_{o}|}$. Thus, if one inserts  $\varphi_{o}$ into (\ref{s}), it is easy to check that the right  potential $U$  is
\begin{equation} \label{po}
U(\varphi_{o})=-4\varphi_{o}^2\ln |\varphi_{o}|.
\end{equation}
Now, one can omit the subscript    $_{o}$ and write the above result for $\varphi$ in general. Note that, such a localized  non-topological solution (\ref{ssw}) is essentially unstable and spontaneously breaks apart.

The most important advantage of the relativistic systems is that, if one can find a  solution at rest, the moving version  can be obtained  easily just by applying  a relativistic boost. In other words, one should replace $x$ and $t$ with $\gamma(x-vt)$ and $\gamma(t-vx)$ respectively, where $\gamma=1/\sqrt{1-v^2}$ and $v$ is any arbitrary velocity ($v<1$).  For example,  the moving version of the special solution (\ref{ssw}) is $\varphi_{v}=\exp(-\gamma^2(x-vt)^2)$.
In general, for a system of the scalar fields $\phi_{i}$ ($i=1,2,\cdots,N$), if one can find a special solution at rest: $\phi_{io}(x,t)$ ($i=1,2,\cdots,N$), the moving version of it would be $\phi_{iv}(x,t)=\phi_{io}(\gamma(x-vt),\gamma(t-vx))$ ($i=1,2,\cdots,N$).  Moreover,  the same standard relativistic  energy ($E$)-rest energy ($E_{o}$)-momentum ($P$) relations  would   exist  between the moving and non-moving versions of any relativistic special solitary wave solution  in general, i.e.  $E_{v}=\gamma E_{o}$ and $P=\gamma E_{o}v$.

Now, instead of the  localized  solution (\ref{ssw}), let us  consider the Lagrangian density (\ref{L}) with a non-localized  un-bounded solution at rest in the following form:
\begin{equation} \label{issw}
\varphi_{o}=\exp(x^2).
\end{equation}
Similar to the same approach which yields   the  appropriate     potential   (\ref{po}) for the requested special solution (\ref{ssw}), here one can  find another  appropriate    potential for the  non-localized solution (\ref{issw}) as well:
\begin{equation} \label{po2}
U(\varphi)=4\varphi^2\ln(|\varphi|).
\end{equation}
The moving version of (\ref{issw}) would be
\begin{equation} \label{isswm}
\varphi_{v}=\exp(\gamma^2(x-vt)^2).
\end{equation}
This non-localized solution has no physical valency. But for the speeds larger than  light, if we take the transformations $x\rightarrow \gamma (x-vt)$ and $t\rightarrow \gamma(t-vx)$ as a general rule,  since $v^2>1$ and then $\gamma=1/(i\sqrt{v^2-1})$ would be a pure imaginary number. Thus, the moving solution (\ref{isswm})  turns    into
\begin{equation} \label{fg}
\varphi_{v}=\exp\left(-\frac{(x-vt)^2}{v^2-1}\right),
\end{equation}
which is now a real localized  moving solution and can  be  interesting.  Note that, if $\varphi_{v}$ for the  FTL speeds does not turn to a real function, we would not achieve our goal. In fact, we deliberately  choose the proposed non-moving solution   (\ref{isswm}) as a function of $x^2$ for this goal.
 One can simply check that   Eq.~(\ref{fg}) is also a solution of the general dynamical equation (\ref{de}) with the potential (\ref{po2}).
 Therefore,  the existence of  a fully relativistic field system with an FTL speed solitary wave solution (\ref{fg}) is  mathematically possible.

\begin{figure}[ht!]
  \centering
  \includegraphics[width=100mm]{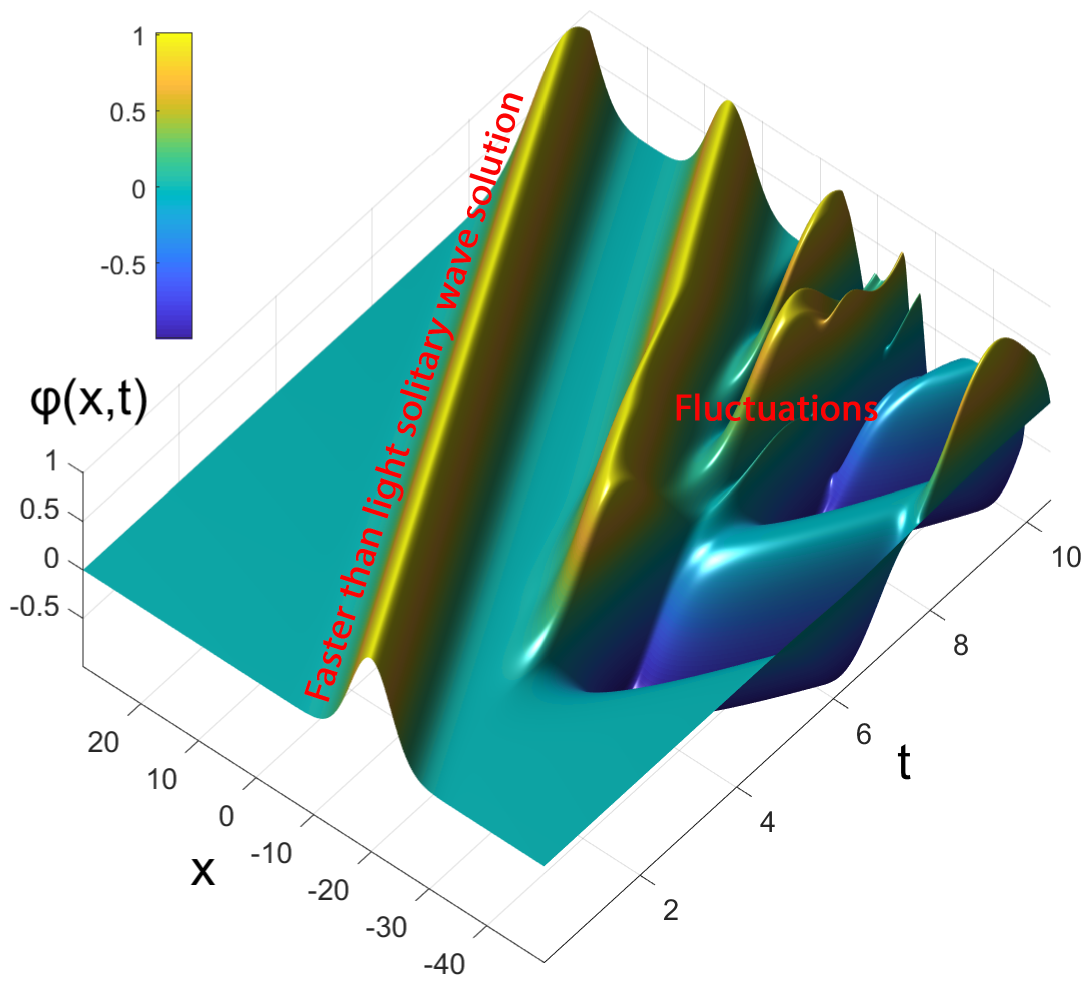}
  \caption{Numerical simulation of the motion of an FTL solitary wave solution (\ref{fg}) of a  real nonlinear KG system (\ref{L}). The   potential (\ref{po2}) in the range $0<\varphi<1$ is negative, then the system and its solutions are essentially  unstable. Hence,  the emergence of some spontaneous  fluctuations is related to this  inherent instability of the system (\ref{L}).} \label{1}
\end{figure}

Numerically or theoretically, it is easy to show that the new   FTL  solitary wave solution  (\ref{fg}) is essentially  unstable. For example,  based on a  finite difference method for the PDE (\ref{de}), one can simply simulate the motion of the  special solitary solution (\ref{fg}) in Matlab for $v=2$. For a brief but remarkable time, it can be seen that a localized solitary wave solution at  an FTL speed can  actually   exist in our simulation program (see Fig.~\ref{1}). After a while,   the form of the  FTL solitary wave solution (\ref{fg}) does not remain stationary and gets disrupted along the time.

In general, according to the Noether's theorem, the  energy density and momentum density, which  belong to the Lagrangian density  (\ref{L}) would be
\begin{equation} \label{df}
\varepsilon(x,t)=\dot{\varphi}^2+\varphi'^2+U(\varphi),\quad \textrm{and} \quad p(x,t)=-2\dot{\varphi}\varphi',
\end{equation}
respectively.  The integration of these functions over  the whole space, for any arbitrary localized solution, yields  the related total energy and total  momentum. Therefore, if one applies these integrations for the FTL  solitary wave solution (\ref{fg}), the following results are obtained:
\begin{equation} \label{ld}
E_{v}=\int_{-\infty}^{\infty}[\dot{\varphi_{v}}^2+\varphi_{v}'^2+
U(\varphi_{v})]dx=\sqrt{\frac{2\pi}{v^2-1}},
\end{equation}
and
\begin{equation} \label{ld2}
 P=\int_{-\infty}^{\infty}[-2\dot{\varphi_{v}}\varphi_{v}']dx=
v\sqrt{\frac{2\pi}{v^2-1}}.
\end{equation}
 Contrary to what we expect, higher  speeds here do not lead to larger total energies. In fact,  a moving solitary wave solution  at the speed of light has  infinite  energy,  while the one moving at $v\rightarrow\infty$ has $E_{v}\rightarrow 0$.
 In regard  to Fig.~\ref{1}, numerical calculations show that despite the instability of the system and the occurrence of  some fluctuations, total  energy and total momentum remain constat according to Eqs.~(\ref{ld}) and (\ref{ld2}) for  case $v=2$, meaning that the energy and momentum conservation laws are valid for such a system as we expected. Moreover, if one expects the same standard  relativistic  relations $E=\gamma E_{o}$, and $P=\gamma E_{o}$ to remain  valid for the FTL solitary wave solution (\ref{fg}), the rest energy must be  an imaginary value $E_{o}=i\sqrt{2\pi}$, resulting from an imaginary  mass.

\section{A  zero-energy soliton solution at FTL speeds}\label{sec3}

In general, for a set of relativistic  scalar  fields $\phi_{i}$ ($i=1,\cdots, N$), the standard Lagrangian densities  are functions of the fields and the kinetic scalars ${\cal S}_{ij}=\partial_{\mu}\phi_{i}\partial^{\mu}\phi_{j}$:
\begin{equation} \label{lags}
	{\cal L}= {\cal L}(\phi_{r},{\cal S}_{ij}),
\end{equation}
where $i,j,r=1,\cdots, N$. According to the principle of least action, the dynamical  equations of motion would be:
\begin{equation} \label{lfg}
	\frac{\partial{\cal L}}{\partial\phi_{i}}-\frac{\partial}{\partial x^{\mu}}\left(\frac{\partial {\cal L}}{\partial \phi_{i,\mu}}\right)=\frac{\partial{\cal L}}{\partial\phi_{i}}-\sum_{j=1}^{N}(1+\delta_{ij})\left[\frac{\partial}{\partial x^{\mu}}\left(\frac{\partial {\cal L}}{\partial {\cal S}_{ij}}\right)\partial^{\mu}\phi_{j}+\frac{\partial{\cal L}}{\partial {\cal S}_{ij}}\partial_{\mu}\partial^{\mu}\phi_{j}\right]=0.
\end{equation}
Since the Lagrangian density (\ref{lags}) is invariant under the infinitesimal space-time translations,  there are four continuity equations $\partial_{\mu}T^{\mu\nu}=0$ and  four conserved quantities $P^{\mu}=\int T^{o\mu}d^{3}\textbf{x}$, where
\begin{equation} \label{egb}
	T^{\mu\nu}=\sum_{i=1}^{N}\frac{\partial{\cal L}}{\partial\phi_{i,\mu}}\frac{\partial\phi_{i}}{\partial x_{\nu}}-{\cal L}g^{\mu\nu}
\end{equation}
is called the energy-momentum tensor and $g^{\mu\nu}$  is   the  Minkowski metric. The   $T^{00}$ component of the energy-momentum tensor is the same as energy density function:
\begin{equation} \label{e5b}
	T^{00}=\varepsilon=\sum_{i=1}^{N}\frac{\partial{\cal L}}{\partial\dot{\phi}_{i}}\dot{\phi}_{i}-{\cal L}=\sum_{j=1}^{N}\sum_{i=1}^{N}\frac{\partial{\cal L}}{\partial{\cal S}_{ij}}\dot{\phi}_{i}\dot{\phi}_{j}(\delta_{ij}+1)-{\cal L}.
\end{equation}

A zero-energy solitary wave solution can be introduced as a special localized solution for which the  energy density function (\ref{e5b}) is zero everywhere.  More precisely, a zero-energy solution is a special solution of  $N$ coupled PDEs (\ref{lfg}) for which $\varepsilon=0$. Condition $\varepsilon=0$ can be interpreted  as a new PDE along with $N$ coupled PDEs (\ref{lfg}). It is mathematically unlikely that $N+1$  coupled  PDEs have a common solution  for $N$ fields. 
However, if the Lagrangian density is such that   it and all its derivatives, i.e. ${\cal L}$, $\frac{\partial{\cal L}}{\partial\phi_{i}}$, $\frac{\partial{\cal L}}{\partial{\cal S}_{ij}}$, and $ \frac{\partial}{\partial x^{\mu}}(\frac{\partial {\cal L}}{\partial {\cal S}_{ij}})$,  become zero simultaneously for a special  solution, thus those $N+1$ PDEs  will  be satisfied  automatically and the special solution would be a zero-energy solution.    
Such a situation is possible only if  the Lagrangian density would be a function of the powers of some   scalar functionals, which are all  zero simultaneously  for a special solution. More precisely, for several scalar fields  $\phi_{i}$ ($i=1,\cdots, N$),  if there are a number of independent scalars  $\mathbb{S}_{j}$ ($j=1,\cdots,m$), which are all  zero simultaneously ($\mathbb{S}_{j}=0$) for a special solution,  the general form of an extended KG Lagrangian density   ($k$-field) with  a zero-energy solution  is:
\begin{eqnarray} \label{sf0}
	{\cal L}=\sum_{n_{1}=0}^{\infty}\sum_{n_{2}=0}^{\infty}\cdots\sum_{n_{m}=0}^{\infty} a({n_{1},\cdots,n_{m}})\mathbb{S}_{1}^{n_{1}}\mathbb{S}_{2}^{n_{2}}\cdots\mathbb{S}_{m}^{n_{m}},
\end{eqnarray}
provided $n_{1}+n_{2}+\cdots+n_{m} \geqslant 3$.    Note that,  scalars  $\mathbb{S}_{j}$ ($j=1,\cdots,m$) and  coefficients $a({n_{1},\cdots,n_{m}})$  can be arbitrary    well-defined functions  of the fields and the kinetic scalars ${\cal S}_{ij}$. For example, for a single scalar field $\phi$, an arbitrary scalar functional  $\mathbb{S}={\cal S}+4\varphi^2\ln |\varphi|$ (${\cal S}=\partial_{\mu}\varphi\partial^{\mu}\varphi$) can be introduced, that    $\varphi=\pm\exp(x^2)$ is a solution for condition $\mathbb{S}=0$. Hence,  this   solution would be a canonical zero-energy  solution for Lagrangian density ${\cal L}=\mathbb{S}^3$ as well. In fact, for ${\cal L}=\mathbb{S}^3$ we have $\frac{\partial{\cal L}}{\partial\phi}=3\mathbb{S}^2\frac{\partial\mathbb{S}}{\partial\phi}$, $\frac{\partial{\cal L}}{\partial{\cal S}}=3\mathbb{S}^2$, and $ \frac{\partial}{\partial x^{\mu}}(\frac{\partial {\cal L}}{\partial {\cal S}})=6\mathbb{S}\frac{\partial \mathbb{S}}{\partial x^{\mu}}$, which obviously are all  zero for condition $\mathbb{S}=0$.


Based on what we have said so far, finding a relativistic  system of fields with a zero-energy solution does not seem difficult. However, if we want to have such a solution that fully satisfies the energetical  stability  considerations, there is not  an easy task ahead of us. A solitary wave solution is energetically stable if any arbitrary variation above its background leads to an increase in the total energy.  Therefore,  the energetical  stability  condition  imposes  serious  restrictions   on  the series  (\ref{sf0}),  which causes it to be turned   to  special  formats.  In this regard,   we  introduce  an   extended KG system with a single   energetically  stable FTL  solitary wave solution in $1+1$ dimensions.  
 For this purpose, three scalar fields $\varphi$, $\theta$ and $\psi$ are used  to introduce  a proper  Lagrangian density. First, we introduce five independent relativistic  functional scalars as  follows:
\begin{eqnarray} \label{sd}
&&\mathbb{S}_{1}=\partial_{\mu}\theta\partial^{\mu}\theta-1, \\&&\label{ll1}
\mathbb{S}_{2}=\partial_{\mu}\varphi\partial^{\mu}\varphi+4\varphi^2\ln |\varphi|, \\&&\label{ll2}
\mathbb{S}_{3}=\partial_{\mu}\varphi\partial^{\mu}\varphi+4\varphi\psi,
\\&&\label{sd3}
\mathbb{S}_{4}=\partial_{\mu}\psi\partial^{\mu}\psi+4\psi^2\ln |\varphi|+8\psi^2+4\psi\varphi,
\\&&\label{sd4}
\mathbb{S}_{5}=\partial_{\mu}\psi\partial^{\mu}\psi+4\varphi^2\ln|\varphi|(\ln|\varphi|+1)^2.
\end{eqnarray}
Apart from  $\mathbb{S}_{1}$, these  scalars are built deliberately in such a way that four  conditions $\mathbb{S}_{i}=0$ ($i=2,3,4,5$), as four   independent PDEs,  have   a unique  common solution at rest  in the following form:
\begin{equation} \label{cs}
\varphi_{o}=\pm\exp(x^2),\quad \psi_{o}=\pm x^2 \exp(x^2).
 \end{equation}
The moving version of this solution would be:
\begin{equation} \label{csm}
\varphi_{v}=\pm\exp(\gamma^2(x-vt)^2),\quad \psi_{v}=\pm\gamma^2(x-vt)^2 \exp(\gamma^2(x-vt)^2).
 \end{equation}
Although such functions  for speeds less than the speed of light are non-localized, but for the FTL speeds $v>1$, they turn to localized functions:
\begin{equation} \label{csu}
\varphi_{v}=\pm\exp\left(\frac{-(x-vt)^2}{v^2-1}\right),\quad \psi_{v}=\frac{\mp (x-vt)^2}{v^2-1} \exp\left(\frac{-(x-vt)^2}{v^2-1}\right),
 \end{equation}
which together can be considered as a localized FTL solution for four independent conditions $\mathbb{S}_{i}=0$ ($i=2,3,4,5$).

Now, we build the  Lagrangian density,
\begin{equation} \label{eL}
{\cal L}=B\sum_{i=1}^5{\cal K}_{i}^3,
 \end{equation}
 where $B$ is a positive constant, and
\begin{eqnarray} \label{jj1}
&&{\cal K}_{1}=h_{1}^2\mathbb{S}_{1},\\&&\label{jj2}
 {\cal K}_{2}=h_{2}^2\mathbb{S}_{1}+\mathbb{S}_{2},\\&&\label{jj3}
{\cal K}_{3}=h_{3}^2\mathbb{S}_{1}+\mathbb{S}_{3},\\&&\label{jj4}
{\cal K}_{4}=h_{4}^2\mathbb{S}_{1}+\mathbb{S}_{4},\\&&\label{jj5}
 {\cal K}_{5}=h_{5}^2\mathbb{S}_{1}+\mathbb{S}_{5},
\end{eqnarray}
\begin{eqnarray} \label{hh}
&&h_{1}=\varphi, \\&&\label{jj1}
h_{2}=\varphi(2\ln |\varphi|+1) \\&&\label{jjf}
h_{3}=2\varphi+\psi \\&&\label{jj6}
h_{4}=2\varphi+5\psi+2\psi\ln|\varphi|\\&&\label{jj9}
h_{5}=\sqrt{2}\varphi(\ln|\varphi|+1)^2.
\end{eqnarray}
Note that, since ${\cal K}_{i}$ ($i=1,2,\cdots,5$) are introduced as five independent  linear combinations of  $\mathbb{S}_{i}$ ($i=1,2,\cdots,5$),  five conditions $\mathbb{S}_{i}=0$ are equivalent to ${\cal K}_{i}=0$ ($i=1,2,\cdots,5$).

Using the  Euler-Lagrange equations (\ref{lfg}) for the new Lagrangian density (\ref{eL}), one can easily obtain the related dynamical equations:
\begin{eqnarray} \label{jkt1}
&&\sum_{i=1}^{5} {\cal K}_{i}\left[2(\partial_{\mu}{\cal K}_{i})   \frac{\partial{\cal K}_{i}}{\partial(\partial_{\mu}\theta)}    +   {\cal K}_{i}\partial_{\mu}\left(\frac{\partial{\cal K}_{i}}{\partial(\partial_{\mu}\theta)}\right)       \right]=0,\\&&\label{jkt2}
\sum_{i=1}^{5} {\cal K}_{i}\left[2(\partial_{\mu}{\cal K}_{i})   \frac{\partial{\cal K}_{i}}{\partial(\partial_{\mu}\varphi)}    +   {\cal K}_{i}\partial_{\mu}\left(\frac{\partial{\cal K}_{i}}{\partial(\partial_{\mu}\varphi)}\right)    -    {\cal K}_{i}\frac{\partial{\cal K}_{i}}{\partial \varphi}   \right]=0,\\&&\label{jkt3}
\sum_{i=1}^{5} {\cal K}_{i}\left[2(\partial_{\mu}{\cal K}_{i})   \frac{\partial{\cal K}_{i}}{\partial(\partial_{\mu}\psi)}    +   {\cal K}_{i}\partial_{\mu}\left(\frac{\partial{\cal K}_{i}}{\partial(\partial_{\mu}\psi)}\right)    -    {\cal K}_{i}\frac{\partial{\cal K}_{i}}{\partial \psi}   \right]=0.
\end{eqnarray}
Moreover, the related energy density function  would be
\begin{eqnarray} \label{ED}
&&\varepsilon(x,t)=\frac{\partial{\cal L}}{\partial\dot{\theta}}\dot{\theta}+\frac{\partial{\cal L}}{\partial\dot{\varphi}}\dot{\varphi}+\frac{\partial{\cal L}}{\partial\dot{\psi}}\dot{\psi}-{\cal L}=\sum_{i=1}^{5}B{\cal K}_{i}^{2}\left[3C_{i}
-{\cal K}_{i}\right]=\sum_{i=1}^{5}\varepsilon_{i},
\end{eqnarray}
which is divided  into five  distinct  parts, in which
\begin{equation}\label{cof}
C_{i}=\dfrac{\partial{\cal K}_{i}}{\partial \dot{\theta}}\dot{\theta}+\dfrac{\partial{\cal K}_{i}}{\partial \dot{\varphi}}\dot{\varphi}+\dfrac{\partial{\cal K}_{i}}{\partial \dot{\psi}}\dot{\psi}=
\begin{cases}
 2h_{1}^2\dot{\theta}^{2} & \text{i=1}
\\
2h_{2}^2\dot{\theta}^{2}+2\dot{\varphi}^2 & \text{i=2}
\\2h_{3}^2\dot{\theta}^{2}+2\dot{\varphi}^2
 & \text{i=3}.
 \\2h_{4}^2\dot{\theta}^{2}+2\dot{\psi}^2
 & \text{i=4}.
 \\2h_{5}^2\dot{\theta}^{2}+2\dot{\psi}^2
 & \text{i=5}.
\end{cases}
\end{equation}
After a straightforward calculation, one can obtain:
 \begin{eqnarray} \label{eis1}
&&\varepsilon_{1}=B{\cal K}_{1}^2[5h_{1}^2\dot{\theta}^2+h_{1}^2\theta'^2+\varphi^2]\geq 0,\\ \label{eis2}&&
\varepsilon_{2}=B{\cal K}_{2}^2[5h_{2}^2\dot{\theta}^2+h_{2}^2\theta'^2+5\dot{\varphi}^2+\varphi'^2+\varphi^2+4(\varphi\ln |\varphi|)^2]\geq 0, \\ \label{eis3}&&
\varepsilon_{3}=B{\cal K}_{3}^2[5h_{3}^2\dot{\theta}^2+h_{3}^2\theta'^2+5\dot{\varphi}^2+\varphi'^2+4\varphi^2+4\psi^2]\geq 0,\\ \label{eis4}&&
\varepsilon_{4}=B{\cal K}_{4}^2[5h_{4}^2\dot{\theta}^2+h_{4}^2\theta'^2+5\dot{\psi}^2+\psi'^2+2\psi^2+4(\varphi+2\psi+\psi\ln |\varphi|)^2]\geq 0,\\ \label{eis5}&&
\varepsilon_{5}=B{\cal K}_{5}^2[5h_{5}^2\dot{\theta}^2+h_{5}^2\theta'^2+5\dot{\psi}^2+\psi'^2+2\varphi^2(1+\ln^2 |\varphi|)(1+\ln |\varphi|)^2]\geq 0.
\end{eqnarray}
 Since all bracket terms $[\cdots]$ in Eqs.~(\ref{jkt1})-(\ref{ED}) and (\ref{eis1})-(\ref{eis5}) are multiplied by  the scalar functionals ${\cal K}_{i}$ or ${\cal K}_{i}^2$ ($i=1,2,\cdots,5$),  any set of functions $\theta$, $\varphi$ and $\psi$ for which  ${\cal K}_{i}=0$ ($\mathbb{S}_{i}=0$) simultaneously, is a special zero-energy solution. As  mentioned before, for  four  conditions  $\mathbb{S}_{i}=0$ ($i=2,3,4,5$), there is a unique   localized common solution (\ref{csu}). However, the condition $\mathbb{S}_{1}=0$, which is in no way related to other conditions $\mathbb{S}_{i}=0$ ($i=2,3,4,5$), has diverging  solutions such as $\theta_{v}=\pm t$, $\theta_{v}=\frac{1}{\sqrt{3}}(2t-x)$, $\theta_{v}=\frac{1}{\sqrt{12}}(4t+2x)$, and so on. 
 Hence, for a zero-energy localized solution, for which ${\cal K}_{i}=0$ ($i=1,2,\cdots,5$), the form of $\varphi$ and $\psi$ are unique (\ref{csu}),  but   $\theta$ can be considered as a free field, provided it satisfies the constraint  $\partial_{\mu}\theta\partial^{\mu}\theta=1$. In fact, the  field $\theta$ is expected  to introduce  a system for which all the terms in the   energy density function will be positive definite.



\begin{figure}[htp]

  \centering

  \label{pphi1}

  \begin{tabular}{cc}

    \includegraphics[width=36mm]{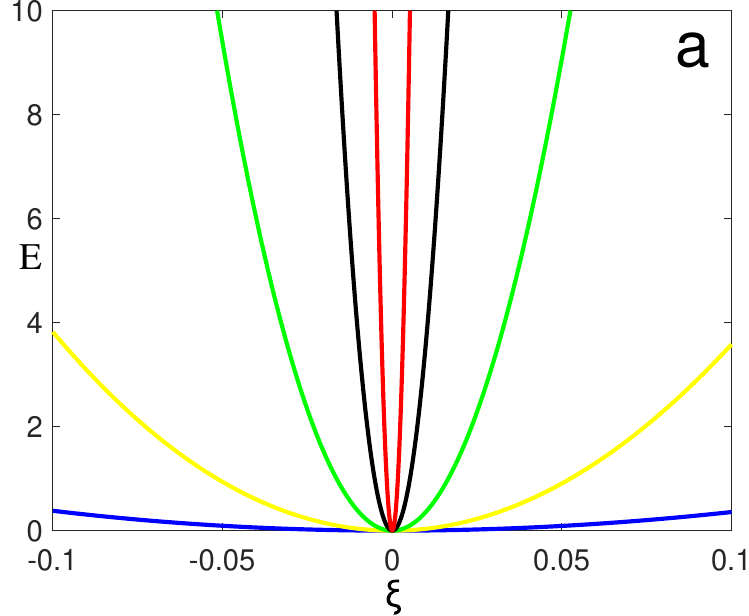}

    \includegraphics[width=36mm]{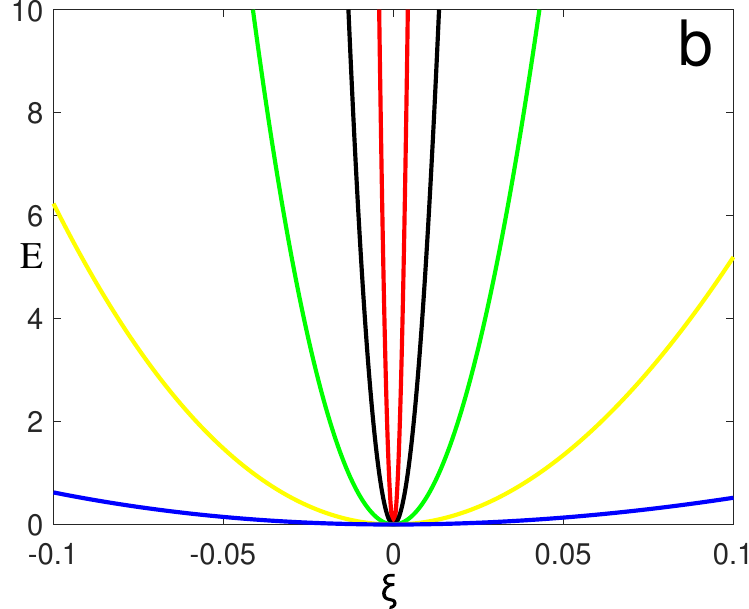}

    \includegraphics[width=36mm]{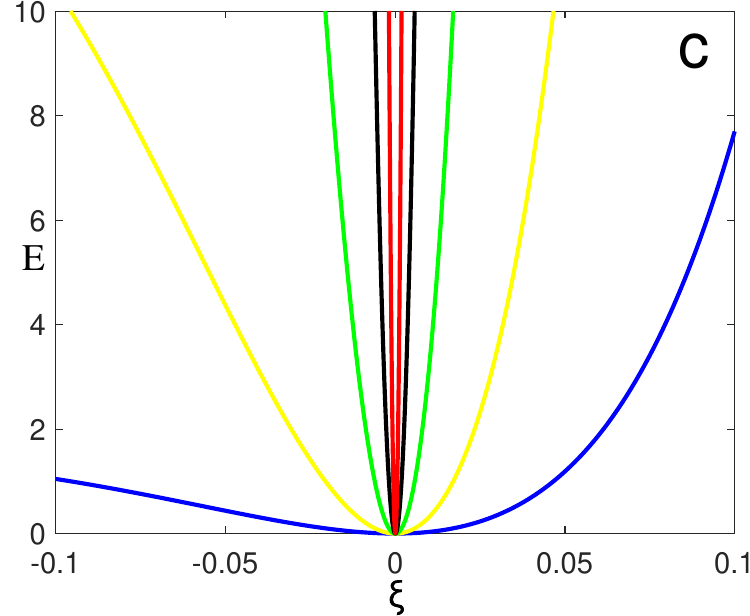}

    \includegraphics[width=36mm]{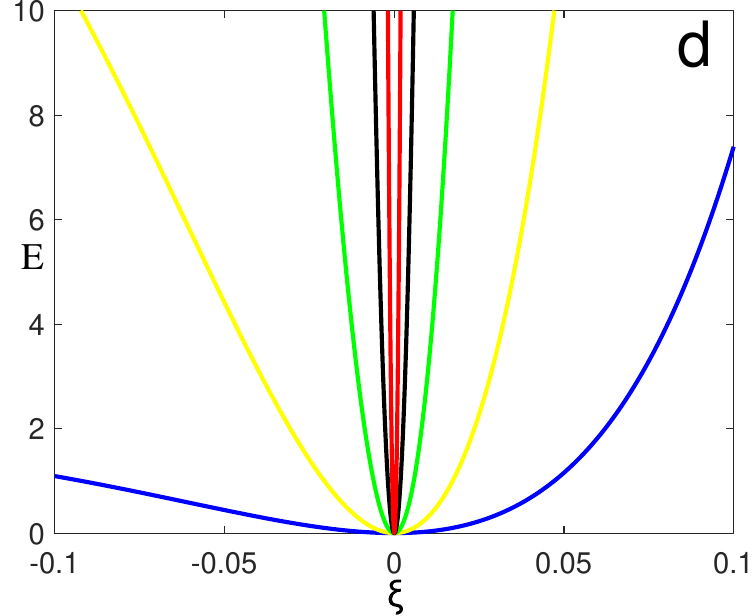}\\

    \includegraphics[width=36mm]{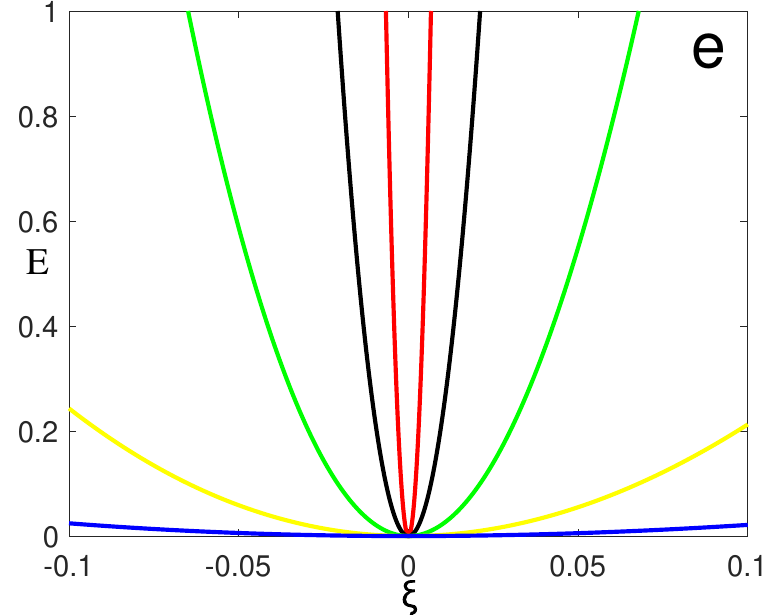}

    \includegraphics[width=36mm]{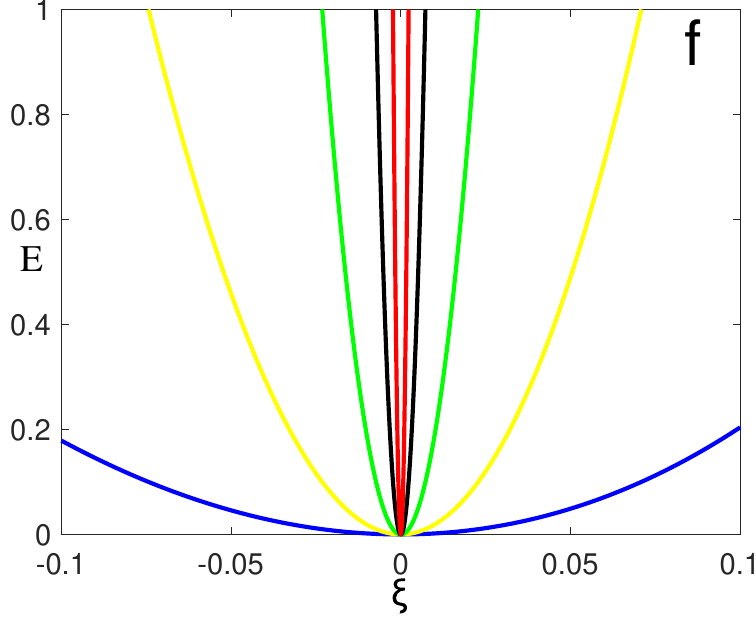}

    \includegraphics[width=36mm]{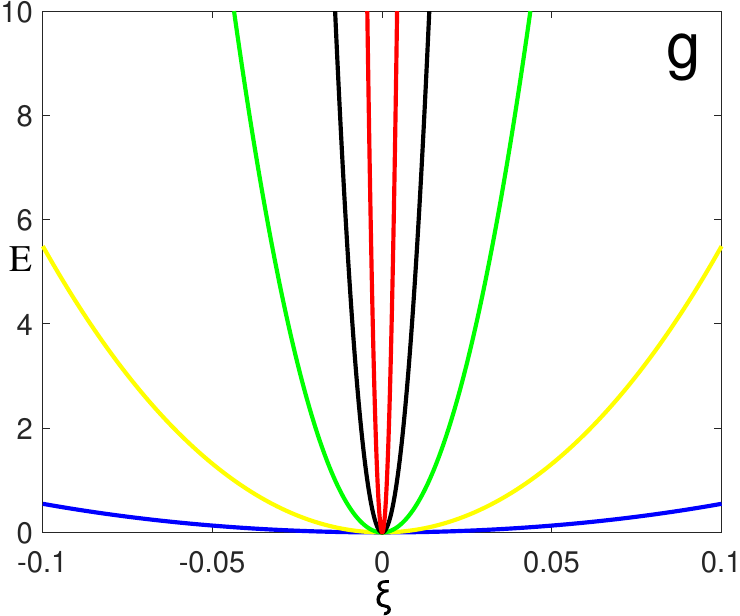}

    \includegraphics[width=36mm]{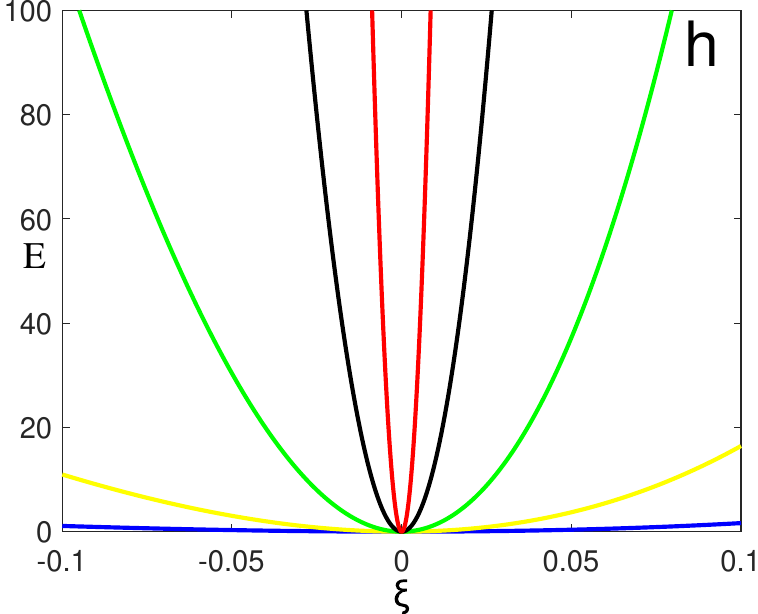}\\

    \includegraphics[width=36mm]{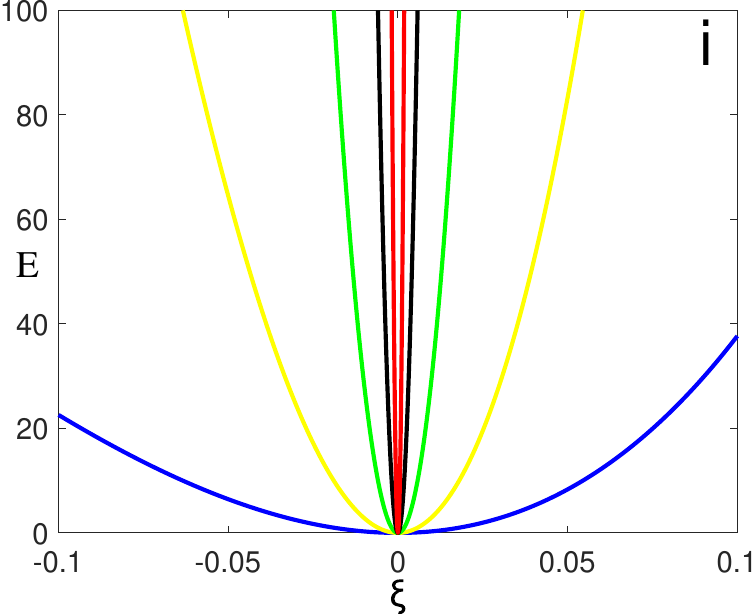}

    \includegraphics[width=36mm]{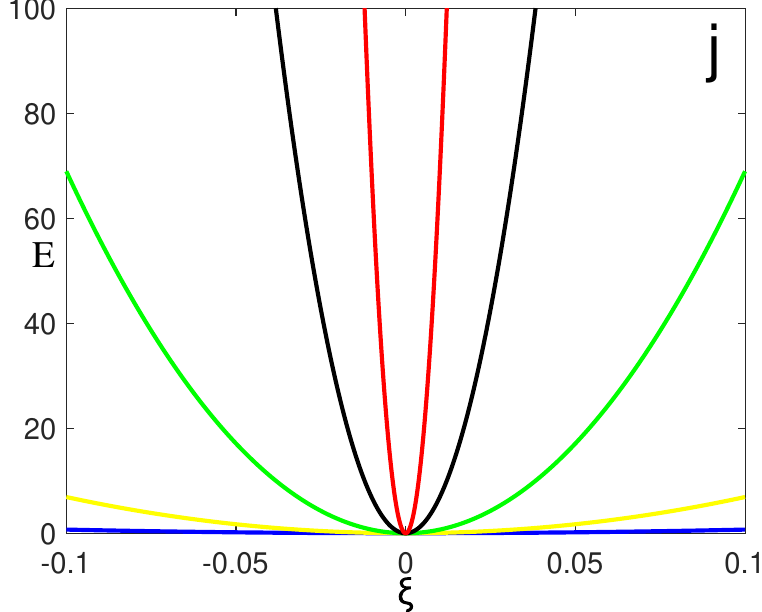}

    \includegraphics[width=36mm]{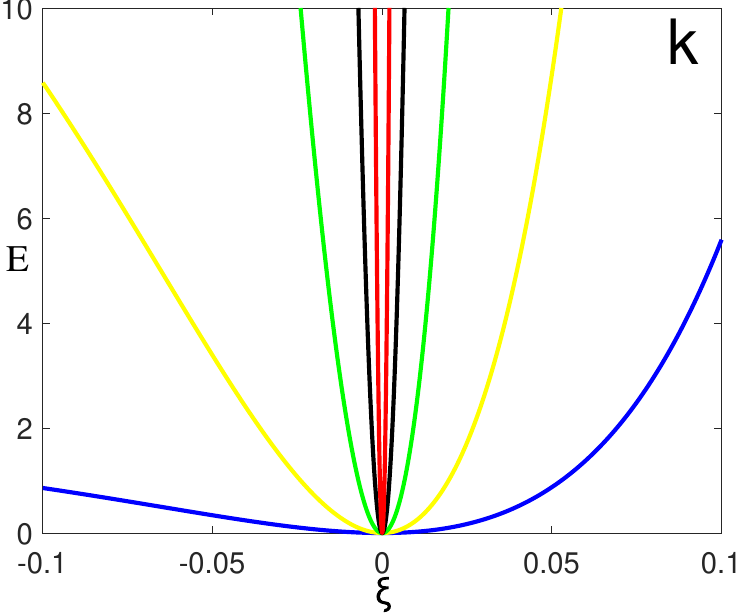}

    \includegraphics[width=36mm]{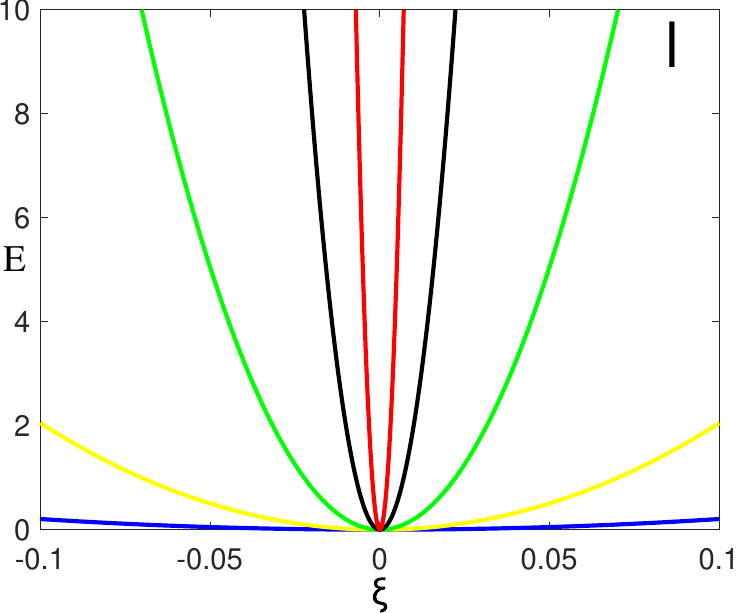}\\

  \end{tabular}
\caption{Plots.~a-l  represent  variations of the total energy $E$ versus small $\xi$ for different deformations  (\ref{var1})-(\ref{var11})  at $t = 0$ and $v=2$ in the context of system (\ref{eL}). Various
colored  curves   blue, yellow, green, black, and red are related to  $B=1$, $B=10$, $B=100$, $B=1000$, and $B=10000$, respectively.}\label{mnb}
\end{figure}

In general, according to Eqs.~(\ref{eis1})-(\ref{eis5}),  all the terms in the energy  density function are positive definite and all ${\cal K}_{i}$'s ($i=1,2,3,4,5$), and subsequently all $\varepsilon_{i}$'s, are zero simultaneously  just for the special solution (\ref{csu}). Therefore,  for any arbitrary  variation above the background of the special solution (\ref{csu}), at least one of the ${\cal K}_{i}$'s ($i=1,2,3,4,5$) would be a non-zero function, and then the energy density function changes to a non-zero positive  function. Thus, for any arbitrary variation, the total energy always increases. In other words, the special solution (\ref{csu}) has the minimum total energy among the other solutions, meaning that it is energetically stable  and then it is a soliton solution.
More precisely, for any  arbitrary (non-trivial) small variations $\delta\varphi$, $\delta\psi$, and $\delta\theta$ above the background of the   special solution (\ref{csu}), i.e. $\varphi=\varphi_{v}+\delta\varphi$, $\psi=\psi_{v}+\delta\psi$, and $\theta=\theta_{v}+\delta\theta$, if one investigates  $\varepsilon_{i}$ ($i=1,2,3,4,5$),  and keep the terms to the least order of variations, it  yields
\begin{eqnarray} \label{so4}
&&\delta\varepsilon_{i}=B[3(C_{i}+\delta C_{i})({\cal K}_{i}+\delta{\cal K}_{i})^{2}-({\cal K}_{i}+\delta{\cal K}_{i})^{3}]=B[3(C_{i}+\delta C_{i})(\delta{\cal K}_{i})^{2}-(\delta{\cal K}_{i})^{3}]\approx\nonumber\\&& \quad\quad B[3C_{i}(\delta{\cal K}_{i})^{2}-(\delta{\cal K}_{i})^{3}]\approx B[3C_{i}(\delta{\cal K}_{i})^{2}]>0,\quad\quad
\end{eqnarray}
where ${\cal K}_{i}=0$ and $\varepsilon_{i}=0$ ($i=1,2,3,4,5$) for the special   solution (\ref{csu}). Therefore, according to Eq.~(\ref{so4}), since $C_{i}>0$,
$\delta\varepsilon_{i}$ ($i=1,2,\cdots,5$), and then $\delta E$, are always positive definite for all small variations, that is, the special  solution (\ref{csu}) is energetically stable.
It should be noted again that, since ${\cal K}_{i}$'s ($i=1,2,\cdots,5$) are five  completely independent functionals  of three scalar fields $\theta$, $\varphi$ and $\psi$, it is not possible for them to be zero simultaneously except when  the special solution (\ref{csu}) along with one of the solutions of $\mathbb{S}_{1}=0$.
For extra evidence, let us consider  the energy variations  for a number of  arbitrary small deformations above the background of the special  solution (\ref{csu}) numerically. For example,  twelve  arbitrary (ad hoc)   deformations can be  introduced as follows:
\begin{eqnarray} \label{var1}
&&\varphi=\pm\exp\left(\frac{-(1+\xi)\tilde{x}^2}{v^2-1}\right),\quad \psi=\frac{\mp \tilde{x}^2}{v^2-1} \exp\left(\frac{-\tilde{x}^2}{v^2-1}\right),\quad\quad\quad\\  \label{As2} &&
\varphi=\pm\exp\left(\frac{-(1+\xi)\tilde{x}^2}{v^2-1}\right),\quad \psi=\frac{\mp \tilde{x}^2}{v^2-1} \exp\left(\frac{-(1+\xi)\tilde{x}^2}{v^2-1}\right),\\&&\label{var2}
\varphi=\pm(1+\xi)\exp\left(\frac{-\tilde{x}^2}{v^2-1}\right),\quad \psi=\frac{\mp (1+\xi)\tilde{x}^2}{v^2-1} \exp\left(\frac{-\tilde{x}^2}{v^2-1}\right), \\&&\label{var3}
\varphi=\pm(1+\xi)\exp\left(\frac{-\tilde{x}^2}{v^2-1}\right),\quad \psi=\frac{\mp \tilde{x}^2}{v^2-1} \exp\left(\frac{-\tilde{x}^2}{v^2-1}\right),\\&&\label{var4}
\varphi=\pm\exp\left(\frac{-\tilde{x}^2}{v^2-1+\xi}\right),\quad \psi=\frac{\mp \tilde{x}^2}{v^2-1+\xi} \exp\left(\frac{-\tilde{x}^2}{v^2-1+\xi}\right),\\&&\label{var5}
\varphi=\pm\exp\left(\frac{-\tilde{x}^2}{v^2-1}\right),\quad \psi=\frac{\mp (1+\xi)\tilde{x}^2}{v^2-1} \exp\left(\frac{-\tilde{x}^2}{v^2-1}\right),\\&&\label{var6}
\varphi=\pm\exp\left(\frac{-\tilde{x}^2}{v^2-1}\right)+\xi\tilde{x}\exp(-\tilde{x}^2),\quad \psi=\frac{\mp \tilde{x}^2}{v^2-1} \exp\left(\frac{-\tilde{x}^2}{v^2-1}\right),\\&&\label{var7}
\varphi=\pm\exp\left(\frac{-(x-(1+\xi)vt)^2}{v^2-1}\right),\quad \psi=\frac{\mp \tilde{x}^2}{v^2-1} \exp\left(\frac{-(x-(1+\xi)vt)^2}{v^2-1}\right),\quad\quad\\&&\label{var8}
\varphi=\pm\exp\left(\frac{-\tilde{x}^2}{v^2-1}\right),\quad \psi=\frac{\mp \tilde{x}^2}{v^2-1} \exp\left(\frac{-\tilde{x}^2}{v^2-1}\right),\quad \theta=(1+\xi)t\\&&\label{var9}
\varphi=\pm\exp\left(\frac{-\tilde{x}^2}{v^2-1}\right),\quad \psi=\frac{\mp (x-vt+\xi)^2}{v^2-1} \exp\left(\frac{-\tilde{x}^2}{v^2-1}\right),\\&&\label{var10}
\varphi=\pm\exp\left(\frac{-\tilde{x}^2}{v^2-1}\right)+\xi\exp(-x^2),\quad \psi=\frac{\mp \tilde{x}^2}{v^2-1} \exp\left(\frac{-\tilde{x}^2}{v^2-1}\right),\\&& \label{var11} \varphi=\pm\exp\left(\frac{-\tilde{x}^2}{v^2-1}\right),\quad \psi=\frac{\mp \tilde{x}^2}{v^2-1} \exp\left(\frac{-\tilde{x}^2}{v^2-1}\right)+\xi t \exp(-\tilde{x}^2).
\end{eqnarray}
where $\tilde{x}=x-vt$, and $\xi$ is a small parameter, which can be considered as an indication of the order of small deformations. The case $\xi = 0$ leads to the same special   solution (\ref{csu}).   For such arbitrary deformations (\ref{var1})-(\ref{var10}) at $t=0$ and $v=2$,  Fig.~\ref{mnb} demonstrates that a larger deformation leads  to a further  increase  in the total energy, as we expected. Except for Eq.~(\ref{var8}), the catalyzer field $\theta$ is considered to be one of the solutions  of $\mathbb{S}_{1}=0$ for all arbitrary  deformations. Furthermore, it is obvious that
  parameter $B$  has a main role in the stability of the special   solution (\ref{csu}),  and its larger values lead to more stability of the special  solution (\ref{csu}).  To put it differently, the larger the values,  the greater will be the increase in the total energy for any arbitrary small variation above the background of the special   solution (\ref{csu}).


For the vacuum solution, i.e. $\varphi=\psi=0$,  according to Eqs.~(\ref{eis1})-(\ref{eis5}), since $h_{i}$'s$=0$ ($i=1,2,\cdots,5$),  all $\varepsilon_{i}$'s$=0$ ($i=1,2,\cdots,5$). This means, when  $\varphi=\psi=0$, it is not at all important  what  the form of the scalar field $\theta$ is.  However, when $\varphi\neq 0$ and $\psi\neq 0$, to have a zero-energy solitary wave solution (\ref{csu}),  $\theta$ must satisfy  the equation  $\partial_{\mu}\theta\partial^{\mu}\theta=1$ (i.e. $\mathbb{S}_{1}=0$). In summary,  the role of the phase  field $\theta$ is like a   path  in all space, along which the zero-energy particle  (\ref{csu}) is stable and moves freely.


\section{A nonzero-energy soliton solution at FTL speeds}

We can now combine the models introduced in sections 2 and 3 and introduce a new model  as follows:
\begin{equation} \label{nmodl}
{\cal L}={\cal L}_{o}+{\cal L}_{cat}=\left[\partial^\mu \varphi\partial_\mu \varphi- U(\varphi)\right]+[B\sum_{i=1}^5{\cal K}_{i}^3],
 \end{equation}
 where ${\cal L}_{o}$ and ${\cal L}_{cat}$ are  the same  Lagrangian densities which are introduced in  Eqs.~(\ref{L})  and (\ref{eL}), respectively. In this model, $U(\varphi)=4\varphi^2\ln(|\varphi|)$, and ${\cal L}_{cat}$  can be called the catalyzer Lagrangian density. For the new system (\ref{nmodl}), the general dynamical equations would be:
 \begin{eqnarray} \label{h1}
&&\sum_{i=1}^{5} {\cal K}_{i}\left[2(\partial_{\mu}{\cal K}_{i})   \frac{\partial{\cal K}_{i}}{\partial(\partial_{\mu}\theta)}    +   {\cal K}_{i}\partial_{\mu}\left(\frac{\partial{\cal K}_{i}}{\partial(\partial_{\mu}\theta)}\right)       \right]=0,\\&&\label{h2}
\left[\Box \varphi +\frac{1}{2}\frac{d U}{d\varphi}\right]+\frac{3}{2}B\sum_{i=1}^{5} {\cal K}_{i}\left[2(\partial_{\mu}{\cal K}_{i})   \frac{\partial{\cal K}_{i}}{\partial(\partial_{\mu}\varphi)}    +   {\cal K}_{i}\partial_{\mu}\left(\frac{\partial{\cal K}_{i}}{\partial(\partial_{\mu}\varphi)}\right)    -    {\cal K}_{i}\frac{\partial{\cal K}_{i}}{\partial \varphi}   \right]=0,\quad\quad\\&&\label{h3}
\sum_{i=1}^{5} {\cal K}_{i}\left[2(\partial_{\mu}{\cal K}_{i})   \frac{\partial{\cal K}_{i}}{\partial(\partial_{\mu}\psi)}    +   {\cal K}_{i}\partial_{\mu}\left(\frac{\partial{\cal K}_{i}}{\partial(\partial_{\mu}\psi)}\right)    -    {\cal K}_{i}\frac{\partial{\cal K}_{i}}{\partial \psi}   \right]=0.
\end{eqnarray}
Moreover, the corresponding energy density function of the new system (\ref{nmodl}) would be:
 \begin{equation} \label{df}
\varepsilon(x,t)=\varepsilon_{o}+\varepsilon_{cat}=[\dot{\varphi}^2+\varphi'^2+U(\varphi)]+\sum_{i=1}^{5}B{\cal K}_{i}^{2}\left[3C_{i}-{\cal K}_{i}\right].
\end{equation}
where $\varepsilon_{o}$ and $\varepsilon_{cat} $  belonging to Lagrangian densities  ${\cal L}_{o}$   and ${\cal L}_{cat}$, respectively.

\begin{figure}[htp]

  \centering

  \label{pphi1}

  \begin{tabular}{cc}

    \includegraphics[width=36mm]{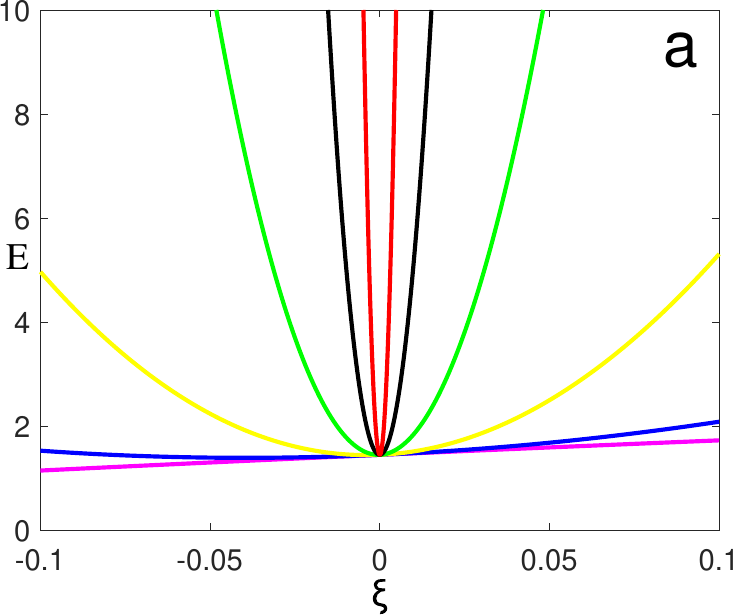}

    \includegraphics[width=36mm]{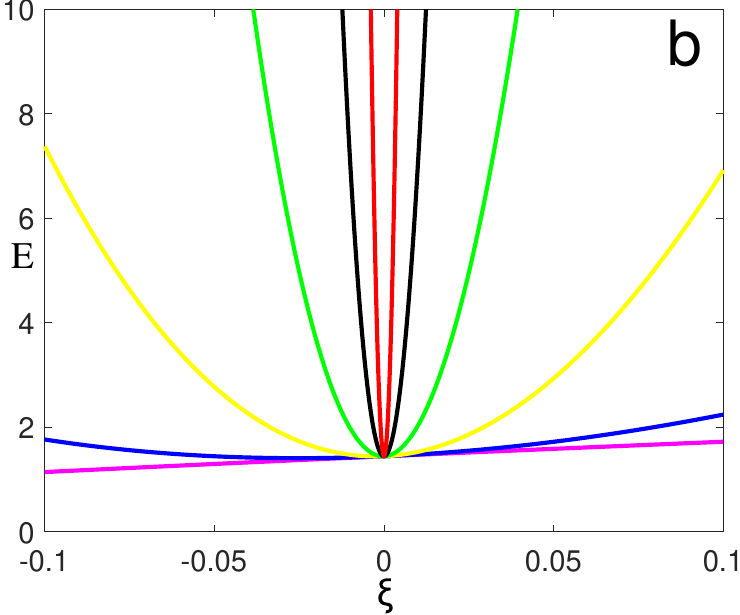}

    \includegraphics[width=36mm]{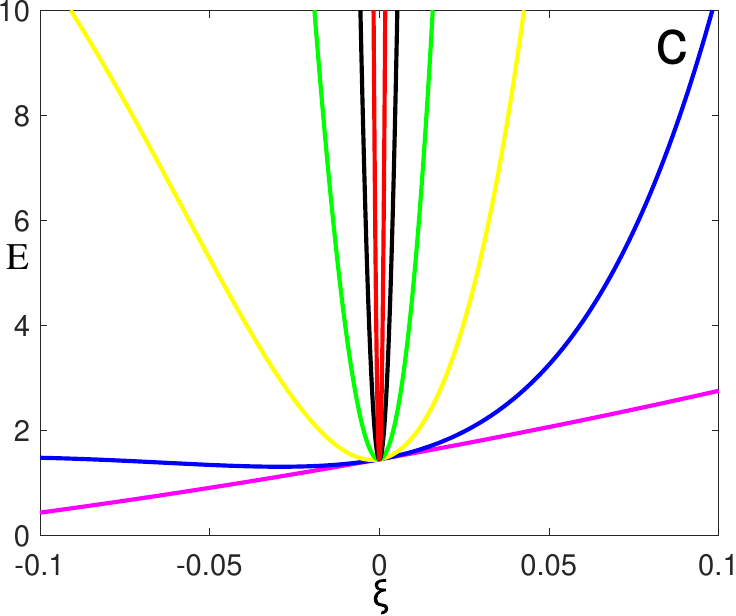}

    \includegraphics[width=36mm]{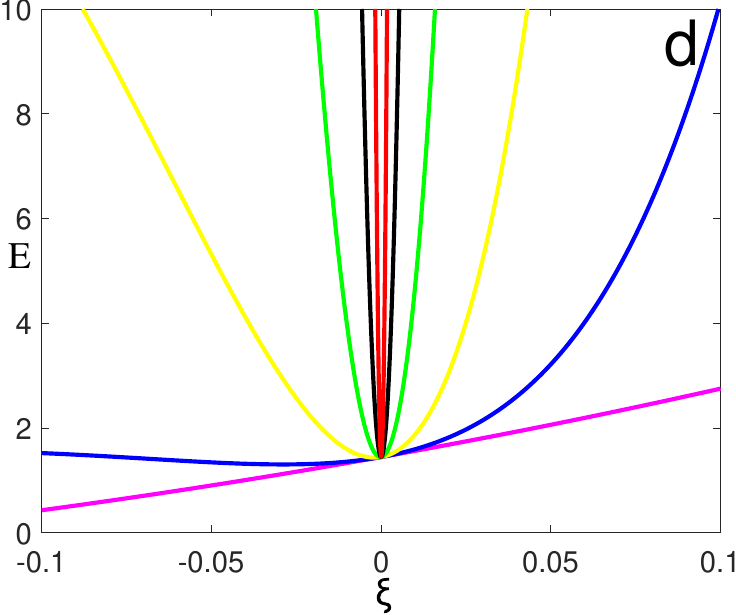}\\

    \includegraphics[width=36mm]{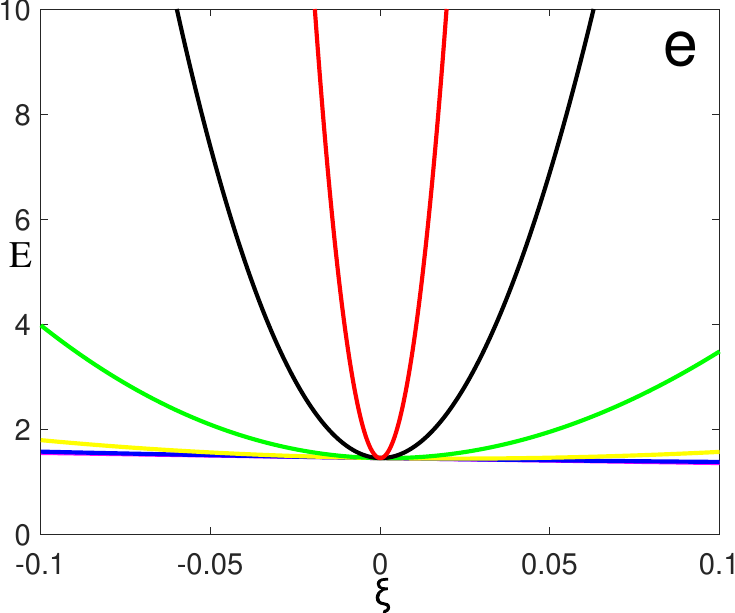}

    \includegraphics[width=36mm]{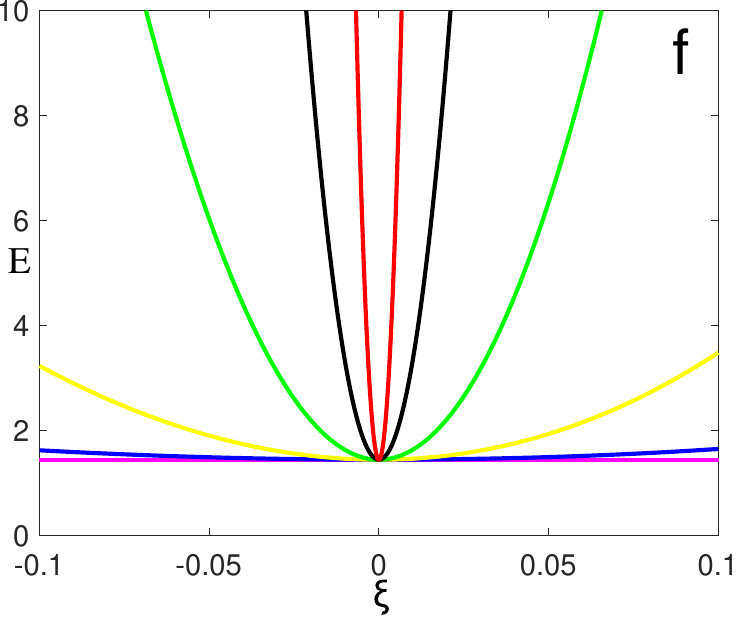}

    \includegraphics[width=36mm]{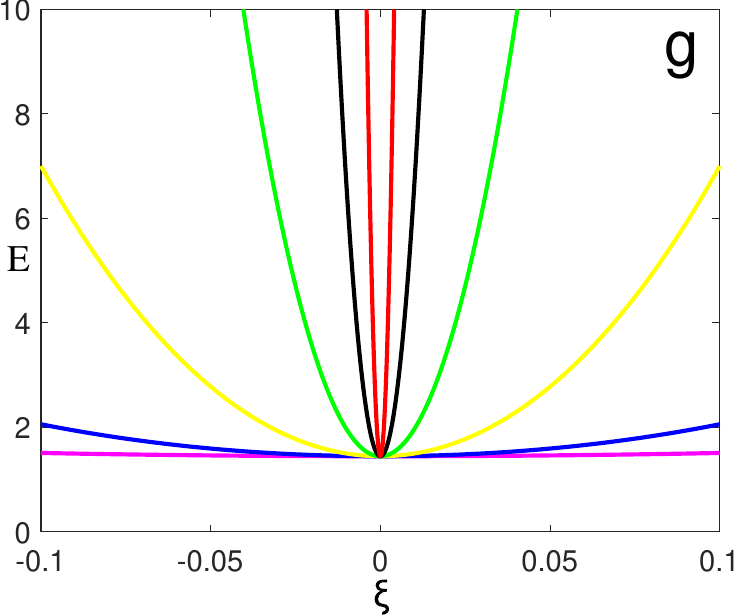}

    \includegraphics[width=36mm]{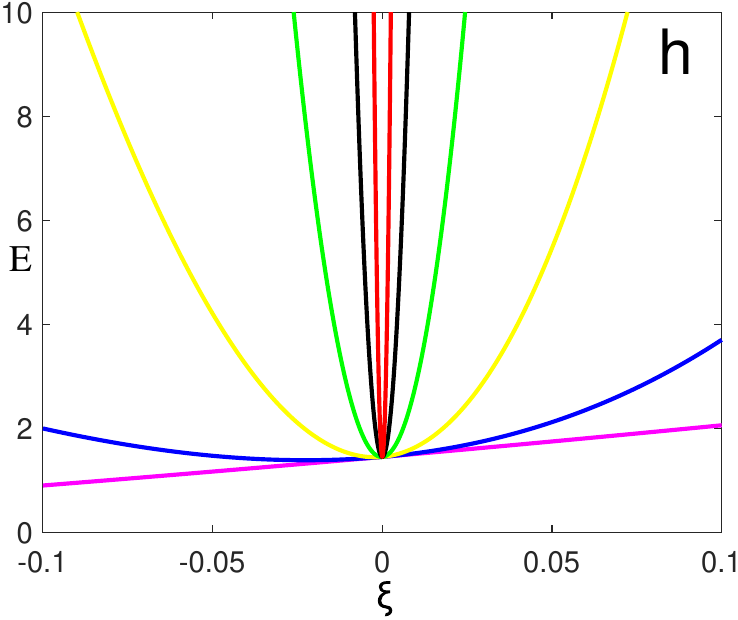}\\

    \includegraphics[width=36mm]{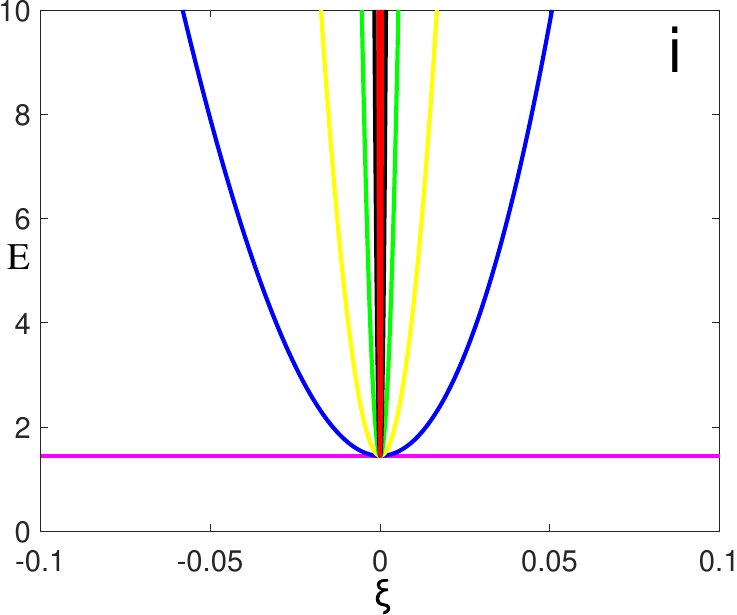}

    \includegraphics[width=36mm]{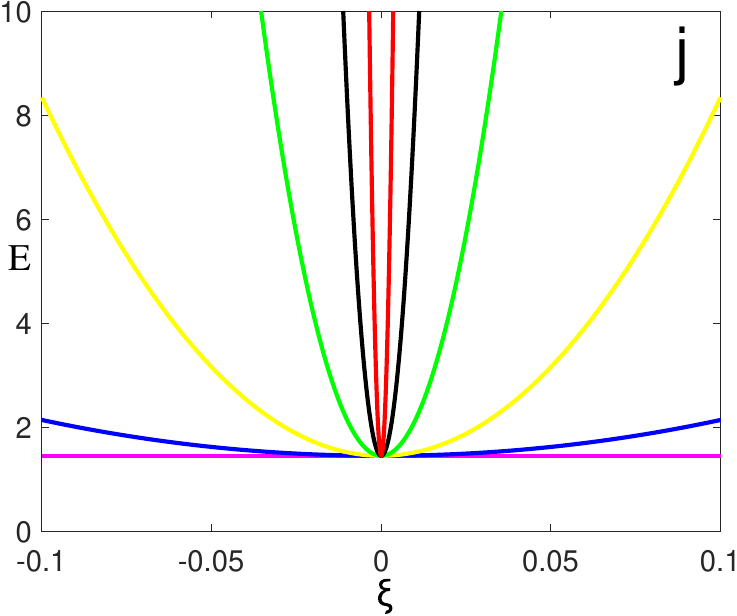}

    \includegraphics[width=36mm]{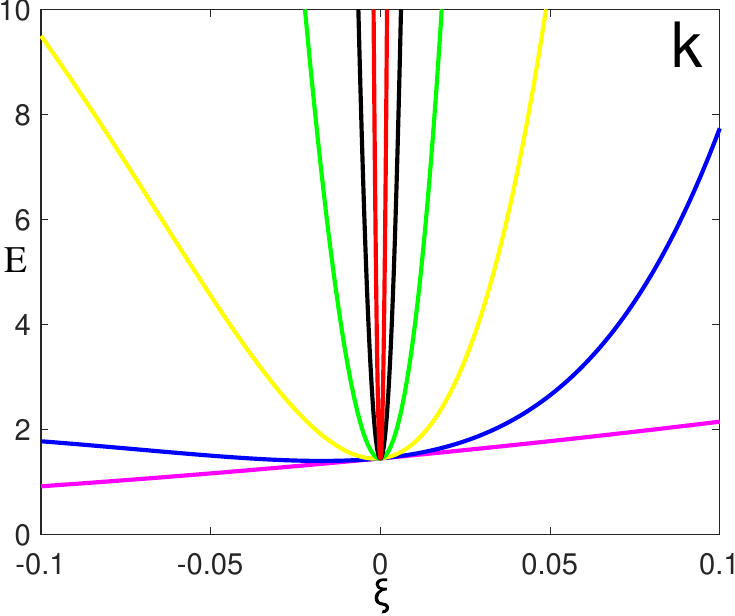}

    \includegraphics[width=36mm]{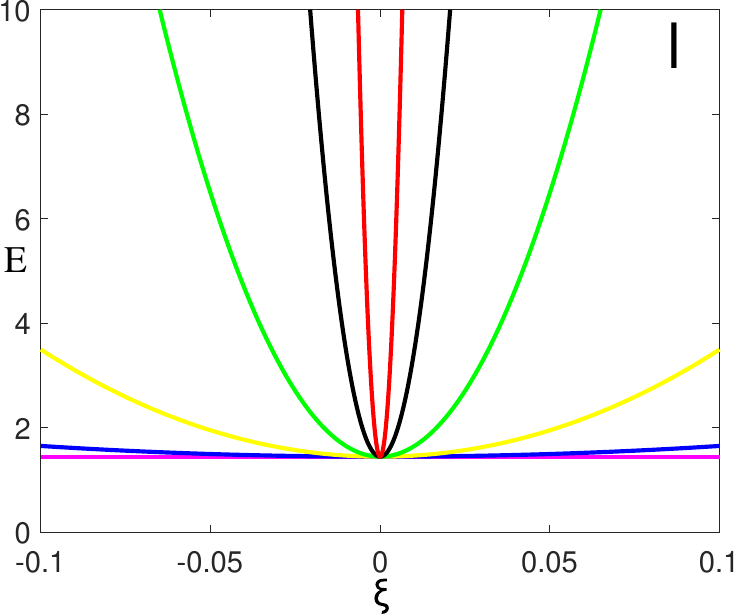}\\

  \end{tabular}
\caption{Plots.~a-l are representing  variations of the total energy $E$ versus small $\xi$ for different deformations  (\ref{var1})-(\ref{var11})  at $t = 0$ and $v=2$ in the context of the new system (\ref{nmodl}). Various
color curves of purple, blue, yellow, green, black, and red are related to $B=0$, $B=1$, $B=10$, $B=100$, $B=1000$, and $B=10000$, respectively. These Figs. confirm how larger values of parameter $B$ lead to more stability. In other words, the
larger the value of $B$, the greater will be the increase in the total energy for any arbitrary small variation
above the background of the FTL soliton solution (\ref{csu}). }\label{dj}
\end{figure}

For the coupled dynamical equations (\ref{h1})-(\ref{h3}), the same Eq.~(\ref{csu}) would again be a solution, but now it is not a zero-energy FTL  solution anymore. In fact, for $\varphi_{v}=\pm\exp\left(-\tilde{x}^2/(v^2-1)\right)$, the expression  $\Box \varphi +\frac{1}{2}\frac{d U}{d\varphi}$ would be independently zero. Furthermore,  ${\cal K}_{i}$'s are all independently zero for the special solution (\ref{csu}) provided $\theta$ is  one of the solutions of $\mathbb{S}_{1}=0$. Hence, all dynamical equations (\ref{h1})-(\ref{h3}) are satisfied automatically for the special solution (\ref{csu}) along with one of the solutions of $\mathbb{S}_{1}=0$. More precisely, just for the  special solution (\ref{csu}), all terms in the dynamical equations, that   contain ${\cal K}_{i}$ and ${\cal K}_{i}^2$, would be automatically zero, and then   the dynamical equations (\ref{h1})-(\ref{h3}) are reduced to   $\Box \varphi +\frac{1}{2}\frac{d U}{d\varphi}=0$ as the  dominant dynamical equation of the   special solution (\ref{csu}).
Since $\varepsilon_{cat}$ is zero for the special solution (\ref{csu}),   the total energy and momentum  as a function of the speed is the same which was obtained  in Eqs.~(\ref{ld}) and (\ref{ld2}),  respectively. In other words, only  the first part of the Lagrangian density  (\ref{nmodl}) is responsible to generate energy and momentum for the special solution (\ref{csu}).


In general,  it can be proved  that the special solution (\ref{csu}) is again  an energetically   stable entity  provided we use a system with a large parameter $B$. For any arbitrary small deformation $\delta\varphi$, $\delta\psi$, and $\delta\theta$ above the background of the special solution (\ref{csu}), which is not now a zero energy FTL soliton in the new model (\ref{nmodl}), the variation of the energy density function would be:
\begin{eqnarray} \label{fhsbn}
&&\delta\varepsilon=\delta\varepsilon_{o}+\delta\varepsilon_{cat}\approx
\left(2\dot{\varphi}_{v}\delta\dot{\varphi}+2\varphi'_{v}\delta\varphi'+
\frac{dU}{d\varphi}\delta\varphi\right)+ \sum_{i=1}^{5}[3BC_{i}(\delta{\cal K}_{i})^{2}].\quad\quad
\end{eqnarray}
The second  term in the energy density variation, i.e. $\delta\varepsilon_{cat}$, is  a positive definite functional of the second order of variations. However, the first term, i.e. $\delta\varepsilon_{o}$, is a functional of the first order of variation $\delta\varphi$ and is not  positive definite. Normally, we expect to exclude the terms that contain second-order of variations  as opposed to the terms that contain first-order of variations. It should be noted that $\delta\varepsilon_{cat}$ contains parameter  $B$, but $\delta\varepsilon_{o}$ does not,  thus, the comparison  between them  needs to be examined more closely. For example, for the case $B=10^{40}$, the inequality $|\delta \varphi|>B(\delta \varphi)^2$ is fulfilled only for the very small variations  less than $10^{-20}$, which are not physically significant. It means that the larger the value of $B$, the smaller variations required  for validation of the inequality $|\delta \varphi|>B(\delta \varphi)^2$.
Likewise, a similar comparison   can be used between  $|\delta\varepsilon_{o}|$ and $\delta\varepsilon_{cat}$. In other words, $|\delta\varepsilon_{o}|$ would be larger than  $\delta\varepsilon_{cat}$ only for the very small variations; meaning that, only for such physically unimportant very small variations,  $\delta\varepsilon=\delta\varepsilon_{o}+\delta\varepsilon_{cat}$ may not be a positive function, and then $\delta E=\int \delta\varepsilon d^3x$ may have a very small negative value (see Fig.~\ref{pphi1}).  Accordingly, for such very small variations $\delta\varphi$, the special solution (\ref{csu})  may not be an energetically  stable entity. However, they are so small physically that  can be ignored in terms of energetical stability criterion.  To summarize, for a large enough value of $B$, $\delta\varepsilon$ is
always positive for all significant physical variations and then the stability of the special solution  (\ref{csu})
would guaranteed appreciably in the context of the new system (\ref{nmodl}). Hence, just for some unimportant too small variations, it may be possible
to see the violation of the energetical  stability criterion, but the  energy reduction for these variations are so small
that they can be ignored physically.

Furthermore, it can be possible to show that  the special solution (\ref{csu}) is really an
energetically stable entity numerically.  For example , we can study numerically the variation of
the total energy for all arbitrary deformations (\ref{var1})-(\ref{var11}) in the context of the new system (\ref{nmodl}) again. Figure~\ref{dj}, which is obtained for $v=2$, demonstrates that for large values of parameter $B$,
the energetical  stability of the special solution (\ref{csu}) would be guaranteed. Similar figures can be obtained for other FTL velocities. In fact, the catalyzer Lagrangian density ${\cal L}_{cat}$ behaves like a massless spook, which surrounds
the special solution (\ref{csu}) and opposes any internal deformation. Although, the   catalyzer  term  strongly  guarantees   the stability of the special solution (\ref{csu}), but it does not appear in any of the observable, and that is why we call it the stability catalyzer.
In sum, the new model (\ref{nmodl}) with  a large value of parameter $B$, leads to  a nonzero-energy   energetically  stable solitary wave solution (\ref{csu}), which moves at FTL speeds. Moreover, Fig.~\ref{dj} shows  that clearly why the case $B = 0$ leads to an unstable  solitary wave solution. In other words, the case $B = 0$ is the same original nonlinear
KG system (\ref{L})  for which  the  energy of the solitary wave solution (\ref{fg}) is not a minimum against   any arbitrary deformation.

\begin{figure}[ht!]
	\centering
	\includegraphics[width=150mm]{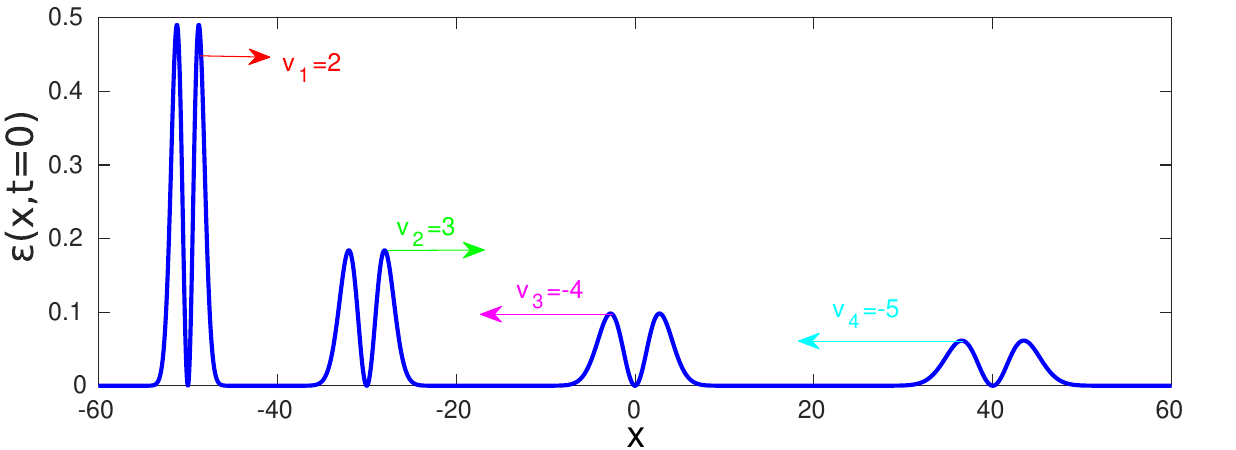}
	\caption{The energy density representation of a four-particle-like solution (\ref{multis}) at $t=0$ in the context of  the new   system (\ref{nmodl}) with $B=10^{10}$. We set the initial velocities and the initial positions   to  $v_{1}=2$, $v_{2}=3$, $v_{3}=-4$, $v_{4}=-5$, $x_{1}=-50$, $x_{2}=-30$, $x_{3}=0$, and $x_{4}=40$, respectively. } \label{aaa}
\end{figure}

Since the special solution (\ref{csu}) is non-topological,  a multi particle-like version of that  can be easily constructed. In general,  the non-topological solutions  are zero at far distances, hence  when they are too far apart, the tail of each non-topological solution would be zero at the position of the others. In other words,  the effect of each non-topological solution on the other solutions is practically zero when the relative distances  between them are large enough.  Therefore, adding any arbitrary  number of the  special solutions (\ref{csu}) together,  when they are initially far enough apart, would still be approximately a solution. In fact, the greater the distances  between the   non-topological solutions, the more accurate this approximation will be.
For example, adding four  distinct  special solutions (\ref{csu}), which initially have different velocities  $v_{i}$ ($i=1,2,3,4$) and stand at different positions $x_{i}$ is  again a solution  at  the initial times (the times that are close to $t=0$):
\begin{eqnarray} \label{multis}
	&& \varphi=\sum_{i=1}^{4}\left[\pm\exp\left(\frac{-(x-v_{i}t-x_{i})^2}{v_{i}^2-1}\right)\right],\nonumber
	\\&& \psi=\sum_{i=1}^{4}\left[\frac{\mp (x-v_{i}t-x_{i})^2}{v_{i}^2-1} \exp\left(\frac{-(x-v_{i}t-x_{i})^2}{v_{i}^2-1}\right)\right],
\end{eqnarray}
provided that $|x_{i+1}-x_{i}|$ be large enough.  For such a linear combination (\ref{multis}), it was observed numerically that the terms  $\mathbb{S}_{i}$ ($i=1,\cdots,5$) are all approximately zero. Hence, based on dynamical equations (\ref{h1})-(\ref{h3}), such a linear combination (\ref{multis}) would be approximately a solution again.   The energy density function of this combination equals  the sum of four  distinct lumps   moving towards each other, as we expected (see Fig.~\ref{aaa}).

 For a multi particle-like solution, the phase field $\theta$ can  change from one solution   to another.    In  regions between two special solutions, the scalar fields  $\varphi$, $\psi$ are zero everywhere.  Thus, there is not any rigorous restriction on $\theta$ to be one of the solutions   of the condition  $\mathbb{S}_{1}=0$. In other words,  where the   scalar fields  $\varphi$, $\psi$ are almost zero, the phase field $\theta$ is completely free and evolves without any rigorous restriction. For example,   if there is a two-particle-like  solution,  the phase field can change  from 
$\theta_{1}=\pm t$  at the position of the first special solution  to $\theta_{2}=\frac{1}{\sqrt{3}}(2t-x)$  at the position of the second one; or it may not change at all and have a definite form, e.g. $\theta=\pm t$,   throughout the space, independent of any number of particle-like solutions.



\section{Summary and Conclusion}

For the relativistic classical field systems in $1+1$ dimensions,   there is a general rule to obtain the form of a moving solution  from its known non-moving  version, by using  $x\rightarrow \gamma (x-vt)$ and $t\rightarrow \gamma(t-vx)$ according  to the  Lorentz transformations, where $\gamma=1/\sqrt{1-v^2}$. Based on this general rule, if one finds  a localized  solution at rest, it turns to a localized   solitary wave solution moving at arbitrary speed ($v<1$). However, for a non-localized solution at rest, which  is a function of   $x^2$, $t^2$, and $xt$, it may turn to a  localized   solitary wave solution at  FTL speeds. In fact, with  speeds faster  than the speed of light, $\gamma$ would be a pure imaginary number, but the  even powers of $\gamma$  would be a real number. Thus, for FTL speeds: $x^2\rightarrow \frac{-1}{v^2-1}(x-vt)^2$, $t^2\rightarrow \frac{-1}{v^2-1}(t-vx)^2$, and $xt\rightarrow \frac{-1}{v^2-1}(x-vt)(t-vx)$ which remain real expressions. For example,  a relativistic field system with a special non-localized solution $\varphi=\exp(x^2)$ at rest,  is not physically interesting since it is not square integrable, but for $v>1$ it turns  to  a localized real function $\varphi=\exp(\frac{-1}{v^2-1}(x-vt)^2)$, which can be physically interesting. Therefore, obtaining a special non-moving solution of a  nonlinear field system may not be of physical importance, but it may turn to  an interesting  localized moving solution at FTL speeds.

In this paper,  we first introduced a standard nonlinear  KG system (\ref{L}) for a single real scalar field $\varphi$, which leads to an unstable nonzero-energy solitary wave solution (\ref{fg}) at FTL speeds.  Thereafter,  an extended  KG system (\ref{eL}) for three scalar fields  $\theta$, $\varphi$ and $\psi$ was  introduced. It was shown that this system has a single non-localized zero-energy solution (\ref{cs}) at rest, but it turns  into a single localized stable zero-energy  solution at FTL speeds (\ref{csu}). In general,  a zero-energy solitary wave solution can be introduced as a special solution whose energy density function is zero everywhere.     All the terms in the   energy density  function of the extended KG system (\ref{eL}) are positive definite and are zero simultaneously  just for the special solution (\ref{csu}). In other words, the special solution  (\ref{csu}) has the minimum energy among the other solutions. In fact, the special solution (\ref{csu}) is an energetically stable entity  for which any arbitrary variation  in its internal structure leads to an increase in the total energy. 
It should be noted that  for  several  scalar fields $\phi_{i}$ ($i=1,\cdots,N$), the relativistic extended KG systems or $k$-fields are  nonstandard field systems which  are not linear in the kinetic scalars  ${\cal S}_{ij}=\partial_{\mu}\phi_{i}\partial^{\mu}\phi_{j}$.



Finally, we showed  how combining   these  systems together  can lead to  a new system (Lagrangian density) with a stable nonzero-energy  solitary wave solution at FTL speeds. Every term in the new lagrangian density (\ref{nmodl}) has now a specific role. The first term, which is the same standard nonlinear KG system (\ref{L}), is responsible to generate energy and momentum for the special solution (\ref{csu}). Although the energy of the second term (\ref{eL}) is zero for the  special solution (\ref{csu}) and it  does not appear in any
of the observable, it is responsible to guarantee the  stability of the special solution (\ref{csu}). Thus, the second term (\ref{eL}) in the new system (\ref{nmodl}) can be called the stability catalyzer.
In fact, the second term (\ref{eL}) behaves like a zero-energy  almost non-deformable backbone
for the particle-like solution (\ref{csu}) at FTL speeds.    The stability catalyzer term contains a parameter $B$, which has a crucial role in the stability. In fact,   the larger the value of $B$, the greater will be the increase in the total energy for
any arbitrary small variation above the background of the special solution (\ref{csu}). Therefore, to be sure that the stability catalyzer term does its role properly, we need to choose a system with a large enough parameter $B$.

It should be noted that although the existence  of a FTL particle is  in direct opposition to observation  and the principle of causality, however    this model, just as an example, shows that  the theory of classical relativistic fields   is not inherently inconsistent with the existence of the  FTL particle-like solitary wave and  soliton solutions in $1+1$ dimensions. It would be interesting to investigate whether FTL models can  also  be built using relativistic fields in   $3+1$ dimensions.


\section*{Acknowledgement}

The author  wishes to express his appreciation to the Persian Gulf University Research Council for their constant support.


\begin{thebibliography}{99}

\bibitem{rajarama} R. Rajaraman, \textit{Solitons and Instantons} (North Holland, Elsevier, Amsterdam, 1982).
\bibitem{Das} A. Das, \textit{Integrable Models} (World Scientific, 1989).
\bibitem{lamb} G. L. Lamb, Jr., \textit{Elements of Soliton Theory} (John Wiley and Sons, USA, 1980).
\bibitem{TS} N. Manton,  P. sutcliffe,  \textit{Topological Solitons}, (Cambridge University Press, 2004).









\bibitem{phi41} Campbell, D. K., Peyrard, M., $\&$ Sodano, P. (1986). \textit{Kink-antikink interactions in the double sine-Gordon equation}. Physica D: Nonlinear Phenomena, \textbf{19(2)}, 165-205.



\bibitem{phi45} Goodman, R. H., $\&$ Haberman, R. (2005). \textit{Kink-Antikink Collisions in the $\phi^4$ Equation: The $n$-Bounce Resonance and the Separatrix Map}. SIAM Journal on Applied Dynamical Systems, \textbf{4(4)}, 1195-1228.

\bibitem{OV} Charkina, O. V., $\&$ Bogdan, M. M. (2006).\textit{ Internal modes of solitons and near-integrable highly-dispersive nonlinear systems}. Symmetry, integrability and geometry: methods and applications, \textbf{2(0)}, 47-12.






\bibitem{Kink1} Dorey, P., Mersh, K., Romanczukiewicz, T., $\&$ Shnir, Y. (2011).   \textit{Kink-antikink collisions in the $\phi^6$  model}. Physical review letters, \textbf{107(9)}, 091602.

\bibitem{Kink2} Gani, V. A., Kudryavtsev, A. E., $\&$ Lizunova, M. A. (2014). \textit{Kink interactions in the $(1+1)$-dimensional $\varphi^6$ model}. Physical Review D, 89(12), 125009.



\bibitem{Kink3} Khare, A., Christov, I. C., $\&$ Saxena, A. (2014). \textit{Successive phase transitions and kink solutions in $\phi^8$, $\phi^{10}$, and $\phi^{12}$ field theories}. Physical Review E, \textbf{90(2)}, 023208.



\bibitem{GMS} Gani, V. A., Moradi Marjaneh, A., $\&$ Saadatmand, D. (2019). \textit{Multi-kink scattering in the double sine-Gordon model}. The European Physical Journal C, \textbf{79(7)}, 620.


\bibitem{Kink5} Bazeia, D., Belendryasova, E., $\&$ Gani, V. A. (2018). \textit{Scattering of kinks of the sinh-deformed $\varphi^ 4$ model}. The European Physical Journal C, \textbf{78(4)}, 340.



\bibitem{Kink6} Gani, V. A., Marjaneh, A. M., Askari, A., Belendryasova, E., $\&$ Saadatmand, D. (2018).\textit{ Scattering of the double sine-Gordon kinks}. The European Physical Journal C, \textbf{78(4)}, 345.




\bibitem{Kink7} Dorey, P., $\&$ Romańczukiewicz, T. (2018). \textit{Resonant kink–antikink scattering through quasinormal modes}. Physics Letters B, \textbf{779}, 117-123.


\bibitem{Kink8} Gani, V. A., Lensky, V., $\&$ Lizunova, M. A. (2015). \textit{Kink excitation spectra in the ($1+ 1$)-dimensional $\varphi^8$ model}. Journal of High Energy Physics, \textbf{2015(8)}, 147.

\bibitem{JRM1} Morris, J. R. (2018). \textit{Small deformations of kinks and walls}. Annals of Physics, \textbf{393}, 122-131.
\bibitem{JRM2} Morris, J. R. (2019). \textit{Interacting kinks and meson mixing}. Annals of Physics, \textbf{400}, 346-365.





\bibitem{Kink10} Bazeia, D., Menezes, R., $\&$ Moreira, D. C. (2018). \textit{Analytical study of kinklike structures with polynomial tails}. Journal of Physics Communications, \textbf{ 2(5)}, 055019.


\bibitem{Kink11}  Christov, I. C., Decker, R. J., Demirkaya, A., Gani, V. A., Kevrekidis, P. G., Khare, A., $\&$ Saxena, A. (2019). \textit{Kink-kink and kink-antikink interactions with long-range tails}. Physical review letters, \textbf{122(17)}, 171601.



\bibitem{Kink12}  Manton, N. S. (2019). \textit{Forces between kinks and antikinks with long-range tails}. Journal of Physics A: Mathematical and Theoretical, \textbf{52(6)}, 065401.

\bibitem{Kink13}   Gani, V. A., Lensky, V., $\&$ Lizunova, M. A. (2015). \textit{Kink excitation spectra in the $(1+ 1)$-dimensional $\varphi^8$ model}. Journal of High Energy Physics, \textbf{2015(8)}, 147.






\bibitem{ana1} Hassanabadi, H., Lu, L., Maghsoodi, E., Liu, G., $\&$ Zarrinkamar, S. (2014). \textit{Scattering of Klein–Gordon particles by a Kink-like potential}. Annals of Physics, \textbf{342}, 264-269.



 
\bibitem{MR1}  Mohammadi, M., $\&$ Riazi, N. (2011). A\textit{pproaching integrability in bi-dimensional nonlinear field equations}. Progress of Theoretical Physics, \textbf{126(2)}, 237-248.

\bibitem{MR3}    Mohammadi, M., $\&$ Riazi, N. (2019). \textit{The affective factors on the uncertainty in the collisions of the soliton solutions of the double field sine-Gordon system}. Communications in Nonlinear Science and Numerical Simulation, \textbf{72}, 176-193.

\bibitem{MR4} Mohammadi, M., $\&$ Dehghani, R. (2021). \textit{Kink-Antikink Collisions in the Periodic $\varphi^4$ Model}. Communications in Nonlinear Science and Numerical Simulation, \textbf{94}, 105575.















\bibitem{waz} Wazwaz, A. M. (2006). \textit{Compactons, solitons and periodic solutions for some forms of nonlinear Klein-Gordon equations}. Chaos, Solitons $\&$ Fractals, \textbf{28(4)}, 1005-1013.




\bibitem{Vak3} Panin, A. G., $\&$ Smolyakov, M. N. (2017). \textit{Problem with classical stability of $U (1)$ gauged Q-balls}. Physical Review D, \textbf{95(6)}, 065006.


\bibitem{Vak4} Kovtun, A., Nugaev, E., $\&$ Shkerin, A. (2018). \textit{Vibrational modes of Q-balls}. Physical Review D, \textbf{98(9)}, 096016.


\bibitem{Vak5} Smolyakov, M. N. (2018). \textit{Perturbations against a Q-ball: Charge, energy, and additivity property}. Physical Review D, \textbf{97(4)}, 045011.



\bibitem{Vak6} Tsumagari, M. I., Copeland, E. J., $\&$ Saffin, P. M. (2008). \textit{Some stationary properties of a Q-ball in arbitrary space dimensions}. Physical Review D, \textbf{78(6)}, 065021






\bibitem{Lee3} Lee, T. D., $\&$ Pang, Y. (1992). \textit{Nontopological solitons}. Physics Reports, \textbf{221(5-6)}, 251-350. 

\bibitem{Scoleman} Coleman, S. (1985). \textit{Q-balls}. Nuclear Physics B, \textbf{262(2)}, 263-283. 

\bibitem{R1} Bazeia, D., Marques, M. A., $\&$ Menezes, R. (2016). \textit{Exact solutions, energy, and charge of stable Q-balls}. The European Physical Journal C, \textbf{76(5)}, 241.

\bibitem{R2} Bazeia, D., Losano, L., Marques, M. A., $\&$ Menezes, R. (2017). \textit{Split Q-balls}. Physics Letters B, \textbf{765}, 359-364.

\bibitem{R3} Anagnostopoulos, K. N., Axenides, M., Floratos, E. G., $\&$ Tetradis, N. (2001). \textit{Large gauged Q balls}. Physical Review D, \textbf{64(12)}, 125006.

\bibitem{R4} Axenides, M., Komineas, S., Perivolaropoulos, L., $\&$ Floratos, M. (2000). \textit{Dynamics of nontopological solitons: Q balls}. Physical Review D, \textbf{61(8)}, 085006.

\bibitem{R5} Bowcock, P., Foster, D., $\&$ Sutcliffe, P. (2009). \textit{Q-balls, integrability and duality}. Journal of Physics A: Mathematical and Theoretical, \textbf{42(8)}, 085403.

\bibitem{R6} Shiromizu, T., Uesugi, T., $\&$ Aoki, M. (1999). \textit{Perturbation analysis of deformed Q-balls and primordial magnetic field}. Physical Review D, \textbf{59(12)}, 125010.

\bibitem{R7} Shiromizu, T. (1998). \textit{Generation of a magnetic field due to excited Q-balls}. Physical Review D, \textbf{58(10)}, 107301.

\bibitem{RM} Ishihara, H., $\&$ Ogawa, T. (2019). \textit{Charge-screened nontopological solitons in a spontaneously broken $U (1)$ gauge theory}. Progress of Theoretical and Experimental Physics, \textbf{2019(2)}, 021B01

\bibitem{R8} Bazeia, D., Losano, L., Marques, M. A., Menezes, R., $\&$ da Rocha, R. (2016). \textit{Compact Q-balls}. Physics Letters B, \textbf{758}, 146-151.





\bibitem{SKrme} Skyrme, T. H. R. (1961). \textit{A non-linear field theory}.   Proceedings of the Royal Society A. \textbf{260}, 127–138.

\bibitem{SKrme2} Skyrme, T. H. R. (1962). \textit{A unified field theory of mesons and baryons}. Nuclear Physics, \textbf{31}, 556-569.


\bibitem{SKrme3} Manton, N. S., Schroers, B. J., $\&$ Singer, M. A. (2004). \textit{The interaction energy of well-separated Skyrme solitons}. Communications in mathematical physics, \textbf{245(1)}, 123-147.

\bibitem{SKrme4} Manton, N. S. (1987). \textit{Geometry of skyrmions}. Communications in Mathematical Physics, \textbf{111(3)}, 469-478.









\bibitem{toft} 't Hooft, G. (1974). \textit{Magnetic monopoles in unified theories}. Nucl. Phys. B, \textbf{79} (CERN-TH-1876), 276-284.

\bibitem{pol} Polyakov, A. M. (1974). \textit{Spectrum of particles in quantum field theory}. JETP Lett, \textbf{20}, 430-433.

\bibitem{MKP} Prasad, M. K. (1980). \textit{Instantons and monopoles in Yang-Mills gauge field theories}. Physica D: Nonlinear Phenomena, \textbf{1(2)}, 167-191.

\bibitem{TOF} Nishino, S., Matsudo, R., Warschinke, M., $\&$ Kondo, K. I. (2018). \textit{Magnetic monopoles in pure Yang-Mills theory with a gauge-invariant mass}. Progress of Theoretical and Experimental Physics, \textbf{2018(10)}, 103B04.

\bibitem{CTOPO}  Eto, M., Hirono, Y., Nitta, M., $\&$ Yasui, S. (2014). \textit{Vortices and other topological solitons in dense quark matter}. Progress of Theoretical and Experimental Physics, \textbf{2014(1)}, 012D01.



















\bibitem{N1} Sassaman, R., $\&$ Biswas, A. (2009). \textit{Topological and non-topological solitons of the generalized Klein–Gordon equations}. Applied Mathematics and Computation, \textbf{215(1)}, 212-220.


\bibitem{N2} Wazwaz, A. M. (2004). \textit{Variants of the generalized KdV equation with compact and noncompact structures}. Computers $\&$ Mathematics with Applications, \textbf{47(4-5)}, 583-591.


\bibitem{N3} Hassan, M. M. (2004). \textit{Exact solitary wave solutions for a generalized KdV–Burgers equation}. Chaos, Solitons $\&$ Fractals, \textbf{19(5)}, 1201-1206.






\bibitem{N4} Cevikel, A. C., Aksoy, E., Güner, Ö., $\&$ Bekir, A. (2013).\textit{Dark-bright soliton solutions for some evolution equations}. Int. J. Nonlinear Sci, \textbf{16}, 195-202.



\bibitem{N5} Lan, Z., $\&$ Gao, B. (2017). \textit{Solitons, breather and bound waves for a generalized higher-order nonlinear Schrödinger equation in an optical fiber or a planar waveguide}. The European Physical Journal Plus, \textbf{132(12)}, 1-13.

\bibitem{N6} Wang, P., Tian, B., Sun, K., $\&$ Qi, F. H. (2015). \textit{Bright and dark soliton solutions and Bäcklund transformation for the Eckhaus–Kundu equation with the cubic–quintic nonlinearity}. Applied Mathematics and Computation, \textbf{251}, 233-242



\bibitem{N7} Jia, T. T., Chai, Y. Z., $\&$ Hao, H. Q. (2017). \textit{Multi-soliton solutions and Breathers for the generalized coupled nonlinear Hirota equations via the Hirota method}. Superlattices and Microstructures, \textbf{105}, 172-182.





\bibitem{N8} Jia, T. T., Gao, Y. T., Feng, Y. J., Hu, L., Su, J. J., Li, L. Q., $\&$ Ding, C. C. (2019). O\textit{n the quintic time-dependent coefficient derivative nonlinear Schrödinger equation in hydrodynamics or fiber optics}. Nonlinear Dynamics, \textbf{96(1)}, 229-241.

\bibitem{N9} Liu, W., Yu, W., Yang, C., Liu, M., Zhang, Y., $\&$ Lei, M. (2017). \textit{Analytic solutions for the generalized complex Ginzburg–Landau equation in fiber lasers}. Nonlinear Dynamics, \textbf{89(4)}, 2933-2939.


\bibitem{N10} Liu, W., Yang, C., Liu, M., Yu, W., Zhang, Y., $\&$  Lei, M. (2017). E\textit{ffect of high-order dispersion on three-soliton interactions for the variable-coefficients Hirota equation}. Physical Review E, \textbf{96(4)}, 042201.

\bibitem{N11} Liu, W., Zhang, Y., Wazwaz, A. M., $\&$ Zhou, Q. (2019). \textit{Analytic study on triple-S, triple-triangle structure interactions for solitons in inhomogeneous multi-mode fiber}. Applied Mathematics and Computation, \textbf{361}, 325-331.






\bibitem{N13} Yan, Y., $\&$ Liu, W. (2019). \textit{Stable transmission of solitons in the complex cubic–quintic Ginzburg–Landau equation with nonlinear gain and higher-order effects}. Applied Mathematics Letters, \textbf{98}, 171-176.










\bibitem{GF} Feinberg, G. (1967). \textit{Possibility of faster-than-light particles}. Physical Review, \textbf{159(5)}, 1089.


\bibitem{GF2} Ehrlich, R. (2003). \textit{Faster-than-light speeds, tachyons, and the possibility of tachyonic neutrinos}. American Journal of Physics, \textbf{71(11)}, 1109-1114.


\bibitem{GF3}  Asaro, C. (1996). \textit{Complex speeds and special relativity}. American Journal of Physics, \textbf{64(4)}, 421-429.

\bibitem{FTL1} Liberati, S., Sonego, S., $\&$ Visser, M. (2002). \textit{Faster-than-c signals, special relativity, and causality}. Annals of physics, \textbf{298(1)}, 167-185.


\bibitem{FTL2} Hill, J. M., $\&$ Cox, B. J. (2012). \textit{Einstein's special relativity beyond the speed of light}. Proceedings of the Royal Society A: Mathematical, Physical and Engineering Sciences, \textbf{468(2148)}, 4174-4192.







\bibitem{Baz1} Bazeia, D., Losano, L., Menezes, R., $\&$ Oliveira, J. C. R. E. (2007). \textit{Generalized global defect solutions}. The European Physical Journal C, \textbf{51(4)}, 953-962.

\bibitem{Adam} Adam, C., Sanchez-Guillen, J., $\&$ Wereszczy\'{n}ski, A. (2007). \textit{$k$-defects as compactons}. Journal of Physics A: Mathematical and Theoretical, \textbf{40(45)}, 13625.

\bibitem{Bab} Babichev, E. (2006). \textit{Global topological $k$-defects}. Physical Review D, \textbf{74(8)}, 085004.



\bibitem{MMQ1}  Mohammadi, M., $\&$ Gheisari, R. (2019). \textit{Zero rest mass soliton solutions}. Physica Scripta, \textbf{95(1)}, 015301.


\bibitem{MMQ2} Mohammadi, M. (2019). \textit{The role of the massless phantom term in the stability of a non-topological soliton solution}. Iranian Journal of Science and Technology, Transactions A: Science, \textbf{43(5)}, 2627-2634.

\bibitem{MMQ4}  Mohammadi, M. (2020). \textit{An energetically stable Q-ball solution in 3+ 1 dimensions}. Physica Scripta, \textbf{95(4)}, 045302.

\bibitem{PH3} Diaz-Alonso, J., $\&$ Rubiera-Garcia, D. (2009). \textit{A study on relativistic lagrangian field theories with non-topological soliton solutions}. Annals of Physics,\textbf{ 324(4)}, 827-873.

\bibitem{MMQ3} Mohammadi, M. (2020). \textit{Stability catalyzer for a relativistic non-topological soliton solution}. Annals of Physics, 168304.





\bibitem{SH} Saha, A., $\&$ Talukdar, B. (2014). \textit{Inverse variational problem for nonstandard Lagrangians}. Reports on Mathematical Physics, \textbf{73(3)}, 299-309.

\bibitem{ZZ} Musielak, Z. E. (2008). \textit{Standard and non-standard Lagrangians for dissipative dynamical systems with variable coefficients}. Journal of Physics A: Mathematical and Theoretical, \textbf{41(5)}, 055205.

\bibitem{Rami0} El-Nabulsi, A. R. (2013). \textit{Non-linear dynamics with non-standard Lagrangians}. Qualitative theory of dynamical systems, \textbf{12(2)}, 273-291..




\bibitem{Rami} El-Nabulsi, R. A. (2015). \textit{Classical string field mechanics with non-standard Lagrangians}. Mathematical Sciences, \textbf{9(3)}, 173-179.

\bibitem{Rami2} El-Nabulsi, R. A. (2014). \textit{Nonlinear integro-differential Einstein’s field equations from nonstandard Lagrangians}. Canadian Journal of Physics, \textbf{92(10)}, 1149-1153.

\bibitem{Rami3} El-Nabulsi, R. A. (2013). \textit{Generalizations of the Klein-Gordon and the Dirac equations from non-standard Lagrangians}. Proceedings of the National Academy of Sciences, India Section A: Physical Sciences, \textbf{83(4)}, 383-387.







\bibitem{Vash}  Vasheghani, A., $\&$ Riazi, N. (1996). \textit{Isovector solitons and Maxwell's equations}. International Journal of Theoretical Physics, \textbf{35(3)}, 587-591.

\bibitem{Mahzon} Mahzoon, M. H., $\&$ Riazi, N. (2007). \textit{Nonlinear electrodynamics and NED-inspired chiral solitons}. International Journal of Theoretical Physics, \textbf{46(4)}, 823-831.


\bibitem{Armend} Armend$\acute{a}$riz-Pic$\acute{o}$n, C., Damour, T., $\&$ Mukhanov, V. I. (1999). \textit{$k$-Inflation}. Physics Letters B, \textbf{458(2-3)}, 209-218.

\bibitem{Chiba} Chiba, T., Okabe, T., $\&$ Yamaguchi, M. (2000).\textit{ Kinetically driven quintessence}. Physical Review D, \textbf{62(2)}, 023511.

\bibitem{Armend2} Armendariz-Picon, C., Mukhanov, V., $\&$ Steinhardt, P. J. (2000). \textit{Dynamical solution to the problem of a small cosmological constant and late-time cosmic acceleration}. Physical Review Letters, \textbf{85(21)}, 4438.









\bibitem{Armend3} Armendariz-Picon, C., $\&$ Lim, E. A. (2005). \textit{Haloes of $k$-essence}. Journal of Cosmology and Astroparticle Physics, \textbf{2005(08)}, 007.





\bibitem{cos1} Dimitrijevic, D. D., $\&$ Milosevic, M. (2012, August).\textit{ About non standard Lagrangians in cosmology}. In AIP Conference Proceedings (Vol. \textbf{1472}, No. 1, pp. 41-46). American Institute of Physics.


\bibitem{cos2} Padmanabhan, T., $\&$  Choudhury, T. R. (2002). \textit{Can the clustered dark matter and the smooth dark energy arise from the same scalar field?}. Physical Review D, \textbf{66(8)}, 081301.



\bibitem{cos3} Renaux-Petel, S., $\&$ Tasinato, G. (2009). \textit{Nonlinear perturbations of cosmological scalar fields with non-standard kinetic terms}. Journal of Cosmology and Astroparticle Physics, \textbf{2009(01)}, 012.








\end{thebibliography}
\end{document}